\documentclass[a4paper,12pt]{article}
\usepackage{a4}

\usepackage{amsbsy}
\usepackage{amssymb}
\usepackage{amsmath}
\usepackage{epsfig}
\usepackage{fullpage}

\begin{document}

\def\CA{{\cal A}}
\def\CB{{\cal B}}
\def\CC{{\cal C}}
\def\CD{{\cal D}}
\def\CE{{\cal E}}
\def\CF{{\cal F}}
\def\CG{{\cal G}}
\def\CH{{\cal H}}
\def\CI{{\cal I}}
\def\CJ{{\cal J}}
\def\CK{{\cal K}}
\def\CL{{\cal L}}
\def\CM{{\cal M}}
\def\CN{{\cal N}}
\def\CO{{\cal O}}
\def\CP{{\cal P}}
\def\CQ{{\cal Q}}
\def\CR{{\cal R}}
\def\CS{{\cal S}}
\def\CT{{\cal T}}
\def\CU{{\cal U}}
\def\CV{{\cal V}}
\def\CW{{\cal W}}
\def\CX{{\cal X}}
\def\CY{{\cal Y}}
\def\CZ{{\cal Z}}

\newcommand{\todo}[1]{{\em \small {#1}}\marginpar{$\Longleftarrow$}}
\newcommand{\labell}[1]{\label{#1}\qquad_{#1}} 
\newcommand{\bbibitem}[1]{\bibitem{#1}\marginpar{#1}}
\newcommand{\llabel}[1]{\label{#1}\marginpar{#1}}

\newcommand{\sphere}[0]{{\rm S}^3}
\newcommand{\su}[0]{{\rm SU(2)}}
\newcommand{\so}[0]{{\rm SO(4)}}
\newcommand{\bK}[0]{{\bf K}}
\newcommand{\bL}[0]{{\bf L}}
\newcommand{\bR}[0]{{\bf R}}
\newcommand{\tK}[0]{\tilde{K}}
\newcommand{\tL}[0]{\bar{L}}
\newcommand{\tR}[0]{\tilde{R}}

\newcommand{\btzm}[0]{BTZ$_{\rm M}$}
\newcommand{\ads}[1]{{\rm AdS}_{#1}}
\newcommand{\ds}[1]{{\rm dS}_{#1}}
\newcommand{\eds}[1]{{\rm EdS}_{#1}}
\newcommand{\sph}[1]{{\rm S}^{#1}}
\newcommand{\gn}[0]{G_N}
\newcommand{\SL}[0]{{\rm SL}(2,R)}
\newcommand{\cosm}[0]{R}
\newcommand{\hdim}[0]{\bar{h}}
\newcommand{\bw}[0]{\bar{w}}
\newcommand{\bz}[0]{\bar{z}}
\newcommand{\be}{\begin{equation}}
\newcommand{\ee}{\end{equation}}
\newcommand{\bea}{\begin{eqnarray}}
\newcommand{\eea}{\end{eqnarray}}
\newcommand{\pat}{\partial}
\newcommand{\lp}{\lambda_+}
\newcommand{\bx}{ {\bf x}}
\newcommand{\bk}{{\bf k}}
\newcommand{\bb}{{\bf b}}
\newcommand{\BB}{{\bf B}}
\newcommand{\tp}{\tilde{\phi}}
\newcommand{\twoa}[4]{\left(\begin{array}{cc} #1 & #2 \\ #3 & #4
\end{array}\right)}
\newcommand{\threea}[9]{\left(\begin{array}{ccc} #1 & #2 & #3 \\ #4 &
#5 & #6 \\ #7 & #8 & #9 \end{array} \right)}
\hyphenation{Min-kow-ski}

\def\apr{\alpha'}
\def\str{{str}}
\def\lstr{\ell_\str}
\def\gstr{g_\str}
\def\Mstr{M_\str}
\def\lpl{\ell_{pl}}
\def\Mpl{M_{pl}}
\def\varep{\varepsilon}
\def\del{\nabla}
\def\grad{\nabla}
\def\perp{\bot}
\def\half{\frac{1}{2}}
\def\p{\partial}
\def\perp{\bot}
\def\eps{\epsilon}

\newcommand{\tr}{\mathrm{tr}}

\def\NPB{{\it Nucl. Phys. }{\bf B}}
\def\PL{{\it Phys. Lett. }}
\def\PRL{{\it Phys. Rev. Lett. }}
\def\PRD{{\it Phys. Rev. }{\bf D}}
\def\CQG{{\it Class. Quantum Grav. }}
\def\JMP{{\it J. Math. Phys. }}
\def\SJNP{{\it Sov. J. Nucl. Phys. }}
\def\SPJ{{\it Sov. Phys. J. }}
\def\JETPL{{\it JETP Lett. }}
\def\TMP{{\it Theor. Math. Phys. }}
\def\IJMPA{{\it Int. J. Mod. Phys. }{\bf A}}
\def\MPL{{\it Mod. Phys. Lett. }}
\def\CMP{{\it Commun. Math. Phys. }}
\def\AP{{\it Ann. Phys. }}
\def\PR{{\it Phys. Rep. }}

\renewcommand{\thepage}{\arabic{page}}
\setcounter{page}{1}

\rightline{hep-th/0304061}
\rightline{ITFA-2003-16}
\vskip 0.75 cm
\renewcommand{\thefootnote}{\fnsymbol{footnote}}
\begin{center}
\Large \bf Nonperturbative Superpotentials and\\
Compactification to Three Dimensions
\end{center}
\vskip 0.75 cm

\centerline{{\bf Rutger Boels,\footnote{rhboels@science.uva.nl} 
Jan de Boer,\footnote{jdeboer@science.uva.nl}
Robert Duivenvoorden\footnote{rjduiven@science.uva.nl}
and
Jeroen Wijnhout\footnote{wijnhout@science.uva.nl}
}}
\vskip .5cm
\centerline{\it Instituut voor Theoretische Fysica,}
\centerline{\it Valckenierstraat 65, 1018XE Amsterdam, The Netherlands}
\vskip .5cm

\setcounter{footnote}{0}
\renewcommand{\thefootnote}{\arabic{footnote}}

\begin{abstract}

We consider four-dimensional ${\cal N}=2$ supersymmetric gauge
theories with gauge group $U(N)$ on $R^3 \times S^1$, in the
presence of a classical superpotential. The low-energy quantum
superpotential is obtained by simply replacing the adjoint scalar
superfield in the classical superpotential by the Lax matrix of
the integrable system that underlies the 4d field theory. We
verify in a number of examples that the vacuum structure obtained
in this way matches precisely that in 4d, although the degrees of
freedom that appear are quite distinct. Several features of 4d
field theories, such as the possibility of lifting vacua from
$U(N)$ to $U(tN)$, become particularly simple in this framework.
It turns out that supersymmetric vacua give rise to a reduction of
the integrable system which contains information about the field
theory but also about the Dijkgraaf-Vafa matrix model. The
relation between the matrix model and the quantum superpotential
on $R^3\times S^1$ appears to involve a novel kind of mirror
symmetry.

\end{abstract}

\newpage

\tableofcontents
\section{Introduction}

The dynamics of supersymmetric gauge theories mimics in many ways that
of ordinary QCD, allowing for instance for confinement and chiral symmetry
breaking. Therefore, if we would have exact control over supersymmetric
theories, we could imagine describing non-supersymmetric gauge theories
as a perturbation away from a supersymmetric point, rather than as a perturbation
around a free field theory. In view of the qualitative similarity of the
low-energy dynamics, such a description could be much better behaved than ordinary
perturbation theory. In addition, there are several indications that
supersymmetry will be restored at sufficiently high energies, and this
provides ample motivation for the study of supersymmetric gauge theories.

As a first step towards the complete understanding of supersymmetric gauge
theories, one would like to understand their vacuum structure. A simple
organizing principle to describe the vacuum structure, low-energy gauge
couplings and correlation functions of chiral operators was given by Dijkgraaf
and Vafa in \cite{dva,dvb,dv}. This organizing principle involves a matrix model
which is obtained from the classical superpotential by reducing it to
its zero modes. The sum over the planar diagrams of the matrix model computes
the quantum superpotential as a function of gluino condensate superfields
$S_i$, one for each semisimple gauge group factor that is left unbroken
by a choice of minimum of the classical superpotential. Though this result
was originally derived using topological string theory, one can give
a proof of it directly in perturbation theory \cite{dv2}, or alternatively
using the Konishi anomaly \cite{cdsw}.

The matrix model, which was originally found for a pure ${\cal N}=2$
theory deformed by a superpotential ${\rm Tr} \,W(\Phi)$,
has been successfully generalized to a variety of other
supersymmetric gauge theories. There are, however, still several
conceptual questions that remain. One of these is whether and why
the gluino condensate superfields play such a distinguished role. It is
true, as has been elaborated in \cite{cdsw}, that they effectively control
the complete chiral ring of the gauge theory, but whether they are also
the sufficient and appropriate degrees of freedom for a full low-energy
effective description remains unknown. The precise reason for the
appearance of the Veneziano-Yankielowicz superpotential
$\sim S \log S - c S$ \cite{vy} remains somewhat mysterious as well,
though in the matrix model it can be interpreted in terms of the measure.
Another question is whether the matrix model approach, or a suitable modification
thereof, is applicable to all possible gauge theories with all possible matter
content. One may also wonder whether the integrable system that underlies
the matrix model has any relation to the integrable system that underlies
$N=2$ gauge theories \cite{is1,is2,is3}.

In an attempt to shed light on some of these issues, we consider in this
paper pure ${\cal N}=2$ theories with superpotential ${\rm Tr} \,W(\Phi)$
compactified on a circle with radius $R$. The integrable system that underlies
the four-dimensional theory becomes much more prominent once the theory
is compactified on a circle, and in addition the low-energy degrees
of freedom are quite distinct. Therefore, we should obtain an interesting
perspective on the matrix model results by compactifying the four-dimensional
theories.

The compactification of the pure ${\cal N}=2$ $U(N)$ theory was discussed in
detail in \cite{sw3}. Recall that in four dimensions there is a $2N$-dimensional
moduli space, and at each point in the moduli space there is an
auxiliary Riemann surface known as the Seiberg-Witten curve 
\cite{sw,sw2,swn1,swn2} (for
a review see e.g. \cite{swrev1,swrev2,swrev3}). The low-energy gauge couplings are
given by the periods of the Seiberg-Witten curve, or in other words by
the complex structure of the Jacobian of the curve, which is a $2N$-dimensional
torus. Once the theory is compactified, the moduli space becomes
$4N$-dimensional. The extra moduli come from the four-dimensional gauge field.
The component of the gauge field along the circle is a scalar from the
three-dimensional point of view, and the remaining 3d gauge field can
be dualized to a second scalar. Thus, each $U(1)$ gauge field gives rise
to two real scalars, that combine into one chiral superfield. The expectation
values of these scalars provide the extra moduli. Since the low-energy
gauge couplings were given by the complex structure of the Jacobian,
it should come as no surprise that the moduli space of the compactified
theory is obtained by adding to each point in
the moduli space of the uncompactified theory
the Jacobian of the associated Seiberg-Witten curve. Altogether the moduli
space is a $4N$-dimensional hyperK\"ahler manifold. According to \cite{sw3},
it has a distinguished complex structure that is independent of the
radius $R$ of the circle. Therefore, after including a superpotential,
the holomorphic data (such as the value of the superpotential and
the vacuum structure) will be independent of the radius $R$. This crucial
fact shows that we should be able to recover the four-dimensional
results directly for finite $R$, without the need to take the limit
$R\rightarrow \infty$.

In several ways, life simplifies in the presence of a circle. First of all, 
the non-perturbative dynamics of the three-dimensional theories we consider here
does not involve fractional
instantons, but only regular 3d instantons (which are monopoles
from the 4d point of view). Second of all, there are no degrees of freedom
that become light anywhere on the moduli space. In four dimensions, there
are massless monopoles and dyons on the moduli space, and these play a
crucial role in the quantum physics. In the presence of a superpotential,
some of these degrees of freedom condense, and this mechanism is responsible
for the existence of supersymmetric vacua \cite{sw,fs,bo}. Once the compact
circle is introduced, there are no such massless degrees of freedom for the theories
we consider, and
we therefore expect that the classical superpotential is actually identical
to the quantum superpotential. The only issue is to find the right variables
in which to express the superpotential. These variables are provided by
the underlying integrable system, which in the case of $U(N)$ is the periodic
Toda chain, and whose phase space coincides with the moduli space of the
unperturbed theory. Indeed, we will find that if the classical superpotential
is ${\rm Tr}\,W(\Phi)$, the quantum superpotential is simply ${\rm Tr}\,W(M)$,
with $M$ the Lax matrix of the integrable system, if the order of $W$ is
not too large. This was first conjectured in \cite{dorey1}, where
it is also shown that this correctly produces several 4d vacua in the
case of a mass deformation of the 4d $N=4$ theory, and it was
explained intuitively in \cite{dorey2}. The validity of this conjecture
for massive (i.e. maximally confining) vacua for theories with higher
order superpotentials was demonstrated in \cite{dorey3}.

In this paper we will analyze the minima and other properties
of the quantum superpotential ${\rm Tr}\,W(M)$ for arbitrary
superpotentials $W$ and gauge groups $U(N)$. We will consider both
massless and massive vacua. We will find that the vacuum structure
is in complete agreement with the results obtained in four dimensions,
in a series of examples. A general proof, using the integrable hierarchy
of the periodic Toda chain, will be described in a separate publication
\cite{toappear}. Several features of the 4d theory have a simple
interpretation on $R^3 \times S^1$. Massive vacua correspond to
simple degenerate Lax matrices, lifting vacua from $U(N)$ to
$U(tN)$ corresponds to applying a simple replica procedure to the
Lax matrix, etc. We also discuss the semiclassical expansion on
$R^3 \times S^1$, and the interpretation of the gluino condensate
superfields. The combination $\sum_i S_i $ has a simple interpretation
as a Lagrange multiplier in three dimensions, but the interpretation
of the individual $S_i$ remains unclear, and seems to involve a new version
of mirror symmetry.

The outline of this paper is as follows.
In section~2 we discuss some background material, such as general
properties of $N=2$ theories in three dimensions, and
the vacuum structure in the absence and presence of a superpotential.
In section~3 we describe the general conjecture and explain qualitatively
why we expect it to be correct. In section~4 we discuss several examples
and show that there is perfect agreement with the results obtained in
four dimensions. In section~5 we discuss the semiclassical expansion,
and in section~6 the relation with the integrable hierarchy of the
periodic Toda chain. Supersymmetric minima determine a specific reduction
of the hierarchy that is remarkably similar to the four-dimensional
field theory expressions, and at the same time remarkably similar to
the integrable system that underlies the Dijkgraaf-Vafa matrix model.
In section~7 we discuss the interpretation of the gluino bilinear superfields
$S_i$, and finally we present some puzzles and open problems.

\section{Field theory background material}

\subsection{${\cal N}=2$ theories in $d=4$}

A pure ${\cal N}=2$ theory with gauge group $U(N)$ in four dimensions
has as bosonic fields a scalar field $\Phi$
transforming in the adjoint representation which is part of an ${\cal N}=1$
hypermultiplet, and a gauge field $A_{\mu}$ which is part of an ${\cal N}=1$
vector multiplet. In a classical minimum of the scalar potential we can always diagonalize
$\Phi$,
\be \label{defphi}
\Phi = \left( \begin{array}{ccc} \phi_1 & & 0 \\
 & \ddots & \\ 0 & & \phi_N \end{array} \right) .
\ee
Such a value of $\Phi$ generically breaks the $U(N)$ gauge group to
$U(1)^N$, but the Weyl group $S_N$ also remains unbroken. Because
of the action of the Weyl group, only symmetric polynomials in the $\phi_i$
are gauge invariant objects, and as a basis of these polynomials we
can choose either ${\rm Tr}(\Phi^i)$ with $i=1,\ldots,N$, or alternatively
the coefficients in the characteristic polynomial,
\bea \label{defP}
P_N(x) & \equiv & \det (x - \Phi) \nonumber \\
 & \equiv & x^N + s_1 x^{N-1} + \ldots s_N .
\eea
Classically, the moduli space is parametrized by the values of the $s_i$, and
generically the gauge group is broken to $U(1)^N$, but at special points
some of the non-abelian gauge symmetry is restored. The quantum theory was
first understood for $SU(2)$ in the famous paper of Seiberg and Witten
\cite{sw}, and later generalized to and studied for many other gauge
theories \cite{sw2,swn1,swn2}. 
For a review see e.g. \cite{swrev1,swrev2,swrev3}. It turns out that the
quantum moduli space is still parametrized by complex coordinates $s_i$,
but there are no longer points where a non-abelian gauge symmetry is
restored. Instead, there are special points in the moduli space where
dyons and/or monopoles can become massless. For each point on the moduli
space one can define an auxiliary curve, the Seiberg-Witten curve,
given by
\be \label{swcurve}
y^2 = P^2_N(x) - 4 \Lambda^{2N}
\ee
whose periods determine the gauge couplings of the low-energy effective
theory. The curve (\ref{swcurve}) describes a double cover of the complex
$x$-plane, with $\Lambda$ the dynamically generated scale of the ${\cal N}=2$
theory. The classical limit corresponds to taking the limit
$\Lambda\rightarrow 0$. The curve (\ref{swcurve}) has a Jacobian $T^{2N}$,
which is a complex torus with period matrix $\tau_{ij} = \int_{B_i} \omega_j$,
where $A_i,B_i$ is a standard basis of one-cycles
on the curve, and $\omega_i$ form basis of holomorphic one-forms normalized
so that $\int_{A_i} \omega_j = \delta_{ij}$.
Massless monopoles and/or dyons appear whenever the curve (\ref{swcurve})
degenerates, i.e. some of the one-cycles collapse.

\subsection{${\cal N}=1$ deformations in $d=4$}

Next, we consider what happens when we add a superpotential
\be
\int d^4 x d^2 \theta {\rm Tr}\, W(\Phi)
\ee
with
\be
W(\Phi) = \sum_{m=0}^n \frac{g_{m+1}}{m+1} \Phi^{m+1}
\ee
some polynomial of the superfield $\Phi$. Classically, the equation
\be \label{wpri}
W'(\Phi) \equiv g_{n+1}(\Phi-a_1)(\Phi-a_2)\ldots(\Phi-a_n) = 0,
\ee
must hold and therefore each of the eigenvalues $\phi_i$ of $\Phi$ in equation (\ref{defphi})
needs to be equal to one of the $a_j$. Thus the moduli space is reduced to
a finite set of points, where at least a $U(1)^N$ gauge symmetry remains
unbroken. More precisely, if $a_i$ is occupied $N_i$ times, in other words
$N_i$ of the $\phi_j$ are equal to $a_i$, then the gauge symmetry is broken
to
\be
U(N) \longrightarrow U(N_1) \times \ldots \times U(N_n) .
\ee

In the quantum theory, the pure ${\cal N}=1$ theories with gauge group
$U(N_i)$ that appear classically confine, a nonzero gluino condensate
$\langle \lambda \lambda \rangle \neq 0$ appears, and the gauge group is
broken to $U(1)$ (for $N_i>0$). One can also show that necessarily some
monopoles/dyons in the ${\cal N}=2$ theory have to condense. Thus the breaking
pattern of the gauge symmetry is
\be \label{gbreak}
U(N) \stackrel{{\rm classical}}{\longrightarrow}
U(N_1) \times \ldots \times U(N_n) \stackrel{{\rm quantum}}{\longrightarrow}
U(1)^k
\ee
where $k$ is the number of $N_i$ that are not equal to zero.

There are several approaches known in the literature to understanding 
the quantum theory. The first studies use Seiberg duality \cite{sd1,sd2}.
The precise structure of the vacua can be obtained 
using a brane construction \cite{bo}, one can use pure field theory
methods \cite{bo,csw,civ}, one can use Calabi-Yau geometry and geometric
engineering \cite{civ}, one can use matrix models \cite{dva,dvb,dv} and one
can use the generalized Konishi anomaly \cite{cdsw}. For our purposes,
we will be mainly interested in a comparison to the vacuum structure as
obtained using field theory methods. The field theory results can be
summarized as follows:
\begin{enumerate}
\item First, express the quantities ${\rm Tr}(\Phi^{m+1})$ in terms of
the $s_i$ that appear in (\ref{defP}).
\item Next, determine the submanifold of the ${\cal N}=2$ theory on
which there are at least $N-k$ mutually local massless monopoles/dyons. On this
submanifold, the Seiberg-Witten curve degenerates and can be written in
the form
\be \label{pcollapse}
y^2 = P_N^2(x)- 4 \Lambda^{2N} = H^2_{N-k}(x) T_{2k}(x)
\ee
for some polynomials $H,T$ of degrees $N-k$, $2k$ respectively. As can
be seen from (\ref{pcollapse}), $N-k$ one-cycles have collapsed.
\item Minimize the classical superpotential, expressed in terms of the $s_i$,
on this submanifold. The resulting extrema are the quantum vacua.
\item One can then show that the quantum vacua are in one-to-one
correspondence with points on the moduli space where in addition to
(\ref{pcollapse}) we also have
\be \label{wcollapse}
G^2_{n-k}(x) T_{2k}(x) = W'(x)^2 + f_{n-1}(x)
\ee
for some polynomials $G,f$ of degrees $n-k,n-1$, and where $T$ is the same
polynomial that appears in (\ref{pcollapse}). Equation (\ref{wcollapse}) can
be viewed as a degeneration of the matrix model curve.
\item Finally, we need to check that the classical limit
of $P_N(x)$ is indeed $\prod_i (x-a_i)^{N_i}$, so that it indeed is a quantum
vacuum corresponding to the appropriate classical vacuum.
\end{enumerate}

For future reference, we also briefly summarize the matrix model approach
of \cite{dv}. Their construction starts with the matrix integral
\be
\int d\Phi e^{-\frac{1}{g^2} {\rm Tr}\, W(\Phi)}.
\ee
The symbol $\Phi$ now denotes an $M\times M$ matrix. Next we consider the planar
diagrams in perturbation theory around a classical minimum where
$U(M)\rightarrow U(M_1) \times \ldots \times U(M_n)$. The free energy, 
that is the sum of the connected planar diagrams, is a function of
the $M_i$ and denoted by ${\cal F}(M_1,\ldots,M_n)$. Next, we replace
$gM_i$ by $S_i$ to construct a function ${\cal F}(S_1,\ldots,S_n)$.
With this definition of ${\cal F}$ the quantum superpotential 
for the ${\cal N}=1$ theory with classical superpotential $W$ in a 
minimum where classically $U(N)$ is broken to $U(N_1)\times \ldots
\times U(N_n)$ is then
\be \label{wdv}
W_{\rm eff}  = \sum_i N_i \frac{\partial {\cal F}}{\partial S_i} + \tau 
  \sum_i S_i  .
\ee
The $S_i$ are superfields whose lowest components are the gaugino condensates
${\rm Tr}_{U(N_i)}(\lambda\lambda)$. This quantum superpotential controls
the complete chiral ring of the ${\cal N}=1$ theory and contains in that
sense more information then the field theory result given above. The
field theory results only describe the minima of (\ref{wdv}) and
are therefore recovered by minimizing (\ref{wdv}) with respect to the $S_i$.

A subtlety in (\ref{wdv}) is the inclusion of the Veneziano-Yankielowicz
superpotential 
\be \label{wvy}
W^{(N_i)}_{VY}(S_i) = S_i \left[ \log \left( \frac{\Lambda^{3N_i}}{S_i^{N_i}}
\right) + N_i \right] 
\ee
for each of the classical unbroken gauge groups $U(N_i)$. 
This can be attributed to the measure in the matrix model.
 For a pure ${\cal N}=1$ superpotential, (\ref{wvy}) is the full
quantum superpotential and its minima are at $S=e^{2\pi i t/N} \Lambda^3$
with $t=0,\ldots,N-1$ and $W_{\rm min}=N\Lambda^3 e^{2\pi i t/N}$.

Having reviewed the situation in four dimensions, we now turn to 
compactifications to three dimensions. 

\subsection{${\cal N}=4$ theories in $d=3$}

Three-dimensional theories with ${\cal N}=4$ can be obtained by dimensionally
reducing four-dimensional theories with ${\cal N}=2$. The structure of the 
Coulomb branch of such theories was studied in detail 
in \cite{sw3}, and they exhibit a
rich set of physical phenomena such as mirror symmetry \cite{3dmirror}.
Under the dimensional
reduction, the four-dimensional vector field $A_{\mu}$ decomposes in one
scalar field $r$ and a three dimensional vector field $A_{\alpha}$. The 
three-dimensional vector can in turn be dualized to a second scalar via
$\partial_{\alpha} \sigma = \epsilon_{\alpha\beta\gamma} F^{\beta\gamma}$.
The two scalars combine into a complex scalar $z$, which is part of a 3d 
hypermultiplet. Thus, in three dimensions, the vector multiplet is dual
to a hypermultiplet. This duality can be performed directly in the
Lagrangian for abelian gauge groups (see e.g. \cite{bho}), but not
for non-abelian groups, similar to what happens with electric-magnetic
duality in four dimensions. If we dimensionally reduce from four to
three dimensions, $\sigma$ will be a periodic variable, but $r$ is
unconstrained. If we instead compactify on a circle to go from four
to three dimensions, both $r$ and $\sigma$ are periodic variables. 

Classically, the complex variable $z$ is $r+i\sigma$, and the action
only depends on $Z+\bar{Z}$, where $Z$ is the superfield with lowest component
$z$. In perturbation theory the action remains a function of $Z+\bar{Z}$ only,
but the relation between the vev of $z$, $r$ and $\sigma$ can become
quite complicated. The periodicity of $r$ and $\sigma$ is therefore not
always manifest in terms of $z$. For example, $z$ could be a coordinate
on a torus in Weierstrass form $y^2 = x^3 + ax + b$, but the fact that
this is a torus is not manifest. Non-perturbatively, the action no
longer needs to be a function of $Z+\bar{Z}$, as instantons generate a 
non-trivial dependence on the zero mode of $\sigma$.

The compactification of a pure ${\cal N}=2$ theory on $R^3\times S^1$
yields a three-dimensional theory with a moduli space which is parametrized
by the vevs of $\phi$ and $z$. The gauge symmetry is broken everywhere
to $U(1)^N$, and both $\phi$ and $z$ are diagonal. The moduli space is
a hyperK\"ahler manifold of dimension $4N$, and contains the moduli
space of the four-dimensional theory that was parametrized only by
$\phi$. What we gain by going to three dimensions are the vevs of $z$,
and these parametrize a $2N$-torus, which can be identified with the
Jacobian of the Seiberg-Witten curve. According to \cite{sw3}, one
of the complex structures of the moduli space is independent of the
radius $R$, and this complex structure is the one that will be relevant
once we break to ${\cal N}=2$ in $d=3$. Since it is independent of
$R$, the vacuum structure we find in three dimensions should be directly
related to the vacuum structure in four dimensions.

\subsection{${\cal N}=2$ deformations in $d=3$}

In this section we consider what happens when we add a superpotential
\be
\int d^3 x d^2 \theta {\rm Tr}\, W(\Phi)
\ee
with
\be
W(\Phi) = \sum_{m=0}^n \frac{g_{m+1}}{m+1} \Phi^{m+1}.
\ee

In four dimensions the gauge symmetry was broken according to
(\ref{gbreak}), and since the complex structure of the moduli
space did not depend on $R$, we expect that for every finite value
of $R$ this remains true. Therefore, we expect that the moduli space
collapses to a finite collection of tori of dimension $2k$.

One of the main questions that we would like to answer in this paper
is whether this is indeed true, and whether this can all be described using
a suitable low-energy effective superpotential that depends on a suitable
set of degrees of freedom. This can indeed be done, but to understand
the result we first need to review some aspects of nonperturbative physics
in three-dimensional gauge theories. 

Non-perturbative physics in three dimensions is due to three-dimensional
instantons, which from the four dimensional point of view are monopoles. 
They are classified by $\pi_2(U(N)/U(1)^N)={\bf Z}^{N-1}$. Indeed, for
each simple root there is a corresponding embedding $SU(2)\subset U(N)$, 
and for each such embedding there is a corresponding elementary monopole.
A general monopole configuration is therefore labeled by a set of integers
$\{n_1,\ldots,n_{N-1}\}$, counting the number of elementary monopole 
constituents. In a pure ${\cal N}=2$ theory in three dimensions,
one can count the number of gaugino zero modes in a general monopole
background using the Callias index theorem \cite{callias}, and one
finds that there are two zero modes only if one $n_i=1$ and all other
$n_j$ vanish. We need two fermionic zero modes in order to get a non-trivial
contribution to the superpotential, and therefore only the single elementary
monopoles contribute to superpotential. Their contribution can be explicitly evaluated and the final result for
the superpotential reads \cite{polyakov,daswadia,affleckharveywitten,katzvafa}
\be \label{w3dq}
W_{\rm quantum} = e^{(Z_1-Z_2)/g_3^2} + e^{(Z_2-Z_3)/g_3^2} + \ldots +
 e^{(Z_{N-1}-Z_{N})/g_3^2}.
\ee
Here, $Z_i$ represents the diagonal entries of $Z$, and in the exponents
one recognizes the simple roots of $U(N)$; $g_3$ is the three-dimensional
Yang-Mills coupling. The result (\ref{w3dq}) is exact, and shows runaway
behavior. In a sense, (\ref{w3dq}) is less subtle than the Veneziano-Yankielowicz
effective superpotential (\ref{wvy}), because (\ref{w3dq}) involves a sum
of ordinary instantons, and fractional instantons play no role. The situation
in the presence of matter is quite a bit more subtle and is discussed in
e.g. \cite{bho,ahiss}. 

\subsection{${\cal N}=2$ deformations on $R^3 \times S^1$}

Now we examine what happens if the $4d$ theory is put on $R^3 \times S^1$.
In addition to the monopoles that contributed to (\ref{w3dq}), there
is one more non-trivial gauge field configuration that contributes,
which is the Kaluza-Klein monopole. This is present due to the existence
of large gauge transformations along the $S^1$ \cite{leeyi}.

The KK monopole adds one extra contribution to (\ref{w3dq}), and it
becomes \cite{sw3,katzvafa}
\be
\label{w3}
W_{\rm quantum}=e^{(Z_1-Z_2)/g_3^2} +\ldots +e^{(Z_{N-1}-Z_{N})/g_3^2} +
e^{-1/Rg_3^2} e^{(Z_N-Z_1)/g_3^2} .
\ee
The three-dimensional answer is recovered by taking $R\rightarrow 0$,
whereas the decompactification limit is $R\rightarrow \infty$ while
keeping the dynamical scale
\be \label{defL}
\Lambda^{3N}\equiv e^{-1/g_4^2} \equiv e^{-1/Rg_3^2}
\ee
fixed.

In order to study the minima of (\ref{w3}), and in order to compare
to the results we will find later, we first introduce a different set
of variables
\bea
y_1 & = & e^{(Z_1-Z_2)/g_3^2} \nonumber \\
& \vdots & \nonumber \\
y_{N-1} & = & e^{(Z_{N-1}-Z_{N})/g_3^2} \nonumber \\
y_0 & = & e^{-1/Rg_3^2} e^{(Z_N-Z_1)/g_3^2} .
\eea
The variables $y_i$ are not unconstrained but obey
\be \label{constr}
\prod_{i=0}^{N-1} y_i = \Lambda^{3N}.
\ee
To impose this constraint, we introduce a Lagrange multiplier field
$L$, and with this field the superpotential (\ref{w3}) can be rewritten
as
\be \label{w3a}
W_{\rm quantum} = y_1 + \ldots + y_{N-1} +y_0 + L \log \left( \frac{\Lambda^{3N}}{
\prod_{i=0}^{N-1} y_i } \right) .
\ee
The minima of (\ref{w3a}) are easily found, they are
\be \label{w3min}
y_{i,{\rm min}} = L_{\rm min} = \Lambda^3 e^{2\pi i t/N}, \qquad
W_{\rm min} = N\Lambda^3 e^{2\pi i t/N}.
\ee
Indeed, the dependence on $R$ has dropped out of (\ref{w3a}) and
(\ref{w3min}), and the results are identical to the results obtained
from the Veneziano-Yankielowicz superpotential (\ref{wvy}). 

Before discussing the general case, we discuss the simplest deformation of 
an ${\cal N}=4$ theory to a ${\cal N}=2$ theory, namely by a mass term
$W(\Phi)= \frac{1}{2} m \Phi^2$. In the presence of such a mass term,
there is a pure ${\cal N}=2$ theory at low energies, whose 
scale $\tilde{\Lambda}$ is related to the high energy ${\cal N}=4$ 
scale $\Lambda$ via scale matching as
\be
\tilde{\Lambda^3} = m \Lambda^2 .
\ee
If we substitute this in (\ref{w3a}), and redefine $y_i \rightarrow m y_i$,
the superpotential becomes
\be \label{w3b}
W = m \sum_{i=0}^{N-1} y_i + L \log \left( \frac{\Lambda^{2N}}{
\prod_{i=0}^{N-1} y_i } \right) .
\ee
The extrema of this superpotential are as before, we merely made a change
of variables. The superpotential (\ref{w3b}) only contains gauge degrees
of freedom, but we could also have chosen to include the diagonal
entries of $\Phi$ in the superpotential. They would simply appear
through an extra mass term,
\be
\label{w3mass}
W = m \sum_{i=0}^{N-1} y_i + L \log \left( \frac{\Lambda^{2N}}{
\prod_{i=0}^{N-1} y_i } \right) + \frac{1}{2} m \sum_{i=0}^{N-1} \phi_i^2.
\ee
The minima of (\ref{w3mass}) are the same as before, as $\phi_i=0$ at the 
extremum. As we will see, (\ref{w3mass}) is literally the superpotential
we obtain if we evaluate $\frac{1}{2} m {\rm Tr} M^2$, with $M$ the Lax
matrix of the periodic Toda chain. In general, we will find a superpotential which is a function
of $y_i$ and $\phi_i$, where the $y_i$ are subject to $\prod y_i = \Lambda^{2N}$.

\section{The proposal}

We now turn to deformations involving a general superpotential
$\int d^4x d^2 \theta {\rm Tr} \, W(\Phi)$ on $R^3 \times S^1$. 
As explained in the introduction, since there are no new massless
degrees of freedom on the moduli space, all we need to do is
to figure out how to write the operators ${\rm Tr}(\Phi^m)$ in terms
of suitable holomorphic variables on the $4N$-dimensional hyperK\"ahler
moduli space of the unperturbed theory. The resulting expression should
be the low-energy effective quantum superpotential. 

To find these variables we need to use the fact that the moduli space
is at the same time the phase space of a (complexified) integrable system. 
This integrable system is the periodic Toda chain, and its relation
to the Seiberg-Witten curve and ${\cal N}=2$ theories in four dimensions
was found in \cite{is1,is2,is3}. The periodic Toda chain is described by
the Hamiltonian 
\be
\label{todaham}
H = \frac{1}{2} \sum_{i=1}^N p_i^2 + \Lambda^2 e^{q_1-q_2} + \ldots +
\Lambda^2 e^{q_{N-1}-q_{N}} + \Lambda^2 e^{q_N-q_1}
\ee
with coordinates $q_i$ and momenta $p_i$, and we introduced a parameter 
$\Lambda^2$
which will later be identified with a field theory scale. 
Notice the similarity of this
expression to that in (\ref{w3mass}). The dynamics described by the
Hamiltonian (\ref{todaham}) is integrable: there exists a Lax operator
$M(p_i,q_i)$, which is an $N\times N$ matrix and a function of coordinates
and momenta, such that the time evolution given by the Hamiltonian
(\ref{todaham}) can equivalently be described by the equation
\be
\label{todaeq1}
\frac{\partial M}{\partial t} = [M,{\cal L}(M)].
\ee
The linear operator ${\cal L}$ will be described in more detail in section~6.
One of the implications of (\ref{todaeq1}) is that the variables
$u_k = \frac{1}{k+1} {\rm Tr}(M^k)$ are constants of the equations of
motion. With further work \cite{olive} one can also show that they
Poisson commute. They provide a complete set of action variables and
generate commuting flows
\be \label{commflow}
\frac{\partial M}{\partial t_k} = \{M,u_k\} = [M,{\cal L}(\frac{1}{k+1} M^k)]
\ee
on the phase space. Thus, the phase space admits action-angle variables, and
the angle variables are linear in the variables $t_k$. The $t_k$ provide
a local set of coordinates on the Jacobian of the Seiberg-Witten curve,
and are closely related to gauge transformations (recall that the angular
coordinates that describe the Jacobian were obtained by dualizing the
gauge field). Therefore, it is in many ways natural to identify the
gauge invariant quantities $\frac{1}{k+1}{\rm Tr}(\Phi^{k+1})$ with the 
conserved quantities $u_k = \frac{1}{k+1} {\rm Tr}(M^k)$ of the dynamical
system. This is also what the original discussion of the role
of the integrable system in ${\cal N}=2$ theories in four dimensions 
implies \cite{is1,is2}. Therefore, we conjecture, following 
\cite{dorey1,dorey2,dorey3} that the quantum superpotential can be obtained from the classical one by the rule
\be
\label{prop}
\int d^4 x d^2 \theta {\rm Tr}\,W(\Phi) \longrightarrow
\int d^4 x d^2 \theta {\rm Tr}\,W(M) ,
\ee
with $M$ the Lax matrix of the integrable system. In order to
compare this to the field theory discussion in section~2, we need
to explain the relation between the coordinates and momenta
in (\ref{todaham}) and the field theory degrees of freedom that appeared
in our discussion in section~2. That can be done by comparing
the Hamiltonian (\ref{todaham}) to the superpotential (\ref{w3mass}).
According to (\ref{prop}), these two should be identified with each other.
Therefore, the momenta $p_i$ of the integrable system should naturally be
identified with the eigenvalues $\phi_i$ of the adjoint superfield $\Phi$
as in (\ref{defphi}). Furthermore, the variables $y_i$ can be identified
with $\Lambda^2 e^{q_i-q_{i+1}}$, where the indices on $q$ are 
identified modulo $N$. 

To make the statement in (\ref{prop}) more explicit, we
need to give the explicit form of the Lax matrix. It is given by
\cite{lax}
\begin{equation} \label{deflax}
M =
 \begin{pmatrix}
  p_1        &   \Lambda^2 e^{q_1-q_2}   &   0    & \dots & z         \\
   1            & p_2  &  \Lambda^2 e^{q_2-q_3}   & \dots & 0         \\
   0            &    1    & p_3 & \dots & 0         \\
   .            &    .    &  .     & \dots & \Lambda^2 e^{q_{N-1}-q_N}   \\
   e^{q_N-q_1} z^{-1} &    .    &  .     &   1   & p_N
 \end{pmatrix} \equiv
 \begin{pmatrix}
  \phi_1        &   y_1   &   0    & \dots & z         \\
   1            & \phi_2  &  y_2   & \dots & 0         \\
   0            &    1    & \phi_3 & \dots & 0         \\
   .            &    .    &  .     & \dots & y_{N-1}   \\
   y_{0} z^{-1} &    .    &  .     &   1   & \phi_N
 \end{pmatrix}.
\ee
The parameter $z$ is a so-called spectral parameter. As long as the superpotential
does not contain higher powers than $N-1$, the superpotential is independent
of $z$. If we compute ${\rm Tr}(M^N)$, we find a $z$-dependent constant
that we simply drop. 
Higher powers can also be taken into account, but we defer that
discussion to section~6. 

Our final proposal is therefore that the quantum superpotential is given by
(\ref{prop}), with $M$ given by (\ref{deflax}). The $y_i$ are constrained
to satisfy $\prod y_i = \Lambda^{2N}$, but we will find it convenient
to keep the $y_i$ unconstrained in the Lax matrix $M$, and to impose the
constraint via a Lagrange multiplier term
\be \label{defmult}
L \log\left( \frac{\Lambda^{2N}}{\prod_{i=0}^{N-1} y_i} \right) 
\ee
in the superpotential.

The spectral curve associated to the Lax matrix $M$ is defined
by the equation 
\be \label{defspec}
\det (x - M) \equiv P_N(x) + (-1)^N(z+\Lambda^{2N} z^{-1}) = 0.
\ee
If we introduce a new variable $y=2 z + (-1)^N P_N(x)$, the spectral
curve becomes
\be
y^2 = P_N(x)^2 - 4 \Lambda^{2N}
\ee
which is exactly the same as the Seiberg-Witten curve
(\ref{swcurve}). Therefore, the quantities ${\rm Tr}(M^{k+1})$ provide
coordinates on the moduli space of the ${\cal N}=2$ theory in four
dimensions, and this provides further motivation for the proposal
(\ref{prop}). 

\subsection{Features of the proposal}

Before working our way through a list of examples, we mention some
general features of the superpotential (\ref{prop}). 

First of all, it is interesting to observe that the quantum superpotential
is simply obtained by substituting the Lax matrix in the classical superpotential.
No matrix models need to be solved. Perhaps this is a manifestation
of the observation that the superpotential of a $d$ dimensional gauge theory 
appears to require solving a $d-4$ dimensional auxiliary theory, as 
suggested in \cite{dv5}. A $-1$-dimensional theory indeed requires
no integrations whatsoever.

Another appealing feature of (\ref{prop}) is that it is polynomial in
$y_i$, and therefore the expansion in powers of $y$ corresponds to a
monopole expansion. In the classical limit where $\Lambda\rightarrow 0$,
all $y_i$ should also be taken to zero, and all non-perturbative
effects disappear. What is left is the classical superpotential expressed
as a function of the $\phi_i$. 

In the periodic Toda chain, the center of mass coordinate $\sum q_i$
does not appear, and as a consequence the variables $y_i$ are
not independent but satisfy $\prod_i y_i = \Lambda^{2N}$. This center
of mass coordinate corresponds to the $U(1)$ factor in $U(N)=U(1)\times SU(N)$, 
and this $U(1)$ is always left unbroken. It accordingly never appears
in the superpotential. The remaining $U(1)^{k-1}$ gauge symmetry that
is left unbroken (see (\ref{gbreak})) will emerge as $k-1$ free parameters
that are left after extremizing the superpotential. In other words, the 
superpotential will not have isolated vacua in general, but will
have moduli spaces of vacua of real dimension $2(k-1)$. It is easy to see
that the flows of the integrable system have to map extrema of the
superpotential into extrema. Therefore what should happen is
that $N-k$ of the flows leave the extremum invariant, while the
remaining $k-1$ flows generate the $k-1$ free complex parameters dual
to the unbroken $U(1)^{k-1}$. The gauge couplings of this unbroken
$U(1)^{k-1}$ can be extracted from the spectral curve, by studying 
the complex structure of the Jacobian of the spectral curve at the
extrema. 

The above picture is based on a comparison to the known results in 
four dimensions (see section~2.2), but we still need to show that this is indeed what happens. Intuitively, what happens is the following. The superpotential is a linear combination of action
variables, and correspondingly generates a flow on the moduli space.
In order for the superpotential to have an extremum, this flow needs
to have a stationary point. Since all flows are linear motions on
the Jacobian of the associated curve, this can only happen if the Jacobian
and Seiberg-Witten curve degenerate. What is not clear is why the existence
of a stationary point of a single flow should automatically 
imply that in fact $N-k$ flows are stationary at this point. Correspondingly,
$N-k$ one-cycles of the Seiberg-Witten curve should collapse, and
the Jacobian should degenerate to a $(k-1)$-dimensional complex torus.

The way in this happens is somewhat mysterious, and is best explained
in terms of the integrable Toda system. The relevant equations are 
summarized in section~6, but we defer a full proof of these statements
to a separate publication \cite{toappear}.

The Lagrange multiplier we introduced in (\ref{defmult}) is not just
there for technical convenience. It turns out that the expectation value
of $L$ in an extremum is identical to the value of the $U(N)$ gluino
bilinear superfield $S$ in the corresponding extremum in the four-dimensional
theory. This will be further discussed in section~7.

A last feature we would like to point out is that the Lax matrix is 
not invariant under general permutations, i.e. under the Weyl group
of $U(N)$. It is only form invariant under a ${\bf Z}_N$ subgroup
of cyclic permutations of the $\phi_i$ and the $y_i$. A choice of
classical vacuum configuration corresponds to choosing each
$\phi_i$ equal to some $a_j$ (the solution of $W'(x)=0$, see
(\ref{wpri})), and all $y_i=0$. It is not guaranteed that all choices
of $\phi_i$ have a corresponding minimum of the quantum superpotential.
Because permutation symmetry is broken, it could be that one
needs to order the $\phi_i$ in a suitable way in order to find a 
minimum of the quantum theory. We will see that this is indeed what
happens. Precisely how one should order the eigenvalues in general
is an interesting open problem that we have not yet been able to
solve.

\section{Examples}

In this section we discuss in detail how our proposal works in various examples. For each of the classical vacua
that gives rise to a symmetry breaking pattern as in (\ref{gbreak}), we expect
to find $\prod_{N_i>0} N_i$ vacua, as each $U(N_i)$ low energy gauge group
has $N_i$ distinct quantum vacua. The various choices of vacua show
up as choices of phases in the solutions below. We will compare the results to
the four dimensional results as described in section~2.2.

\subsection{$U(2)$}

\label{sec:u2}

The first example that we will discuss is $U(2)$ with tree level
superpotential
\be
W=g_1 \tr(\Phi) + \frac{g_2}{2} \tr(\Phi^2) + \frac{g_3}{3}
\tr(\Phi^3) .
\ee
The Lax operator in this case reads
\be
M=\twoa{\phi_1}{y_1+z}{1+y_0 z^{-1}}{\phi_2} .
\ee
The polynomial $P(x)$ appearing in the spectral curve equals
\be \label{u2p} P(x) = (x-\phi_1)(x-\phi_2) - y_0 - y_1 .
\ee
The resulting effective superpotential, using (\ref{prop}), reads
\be
W_{\rm eff} = W(\phi_1) + W(\phi_2) +
(y_0+y_1)(g_2+g_3(\phi_1+\phi_2)) + L
\log\left(\frac{\Lambda^4}{y_0 y_1}\right).
\ee
The equations for the extrema of $W_{\rm eff}$ are

\bea \label{u2a} 
\begin{array}{ccc}
W'(\phi_1) + g_3(y_0+y_1) & = & 0 \nonumber  \\
W'(\phi_2) + g_3(y_0+y_1) & = & 0 \nonumber
\end{array}
& & 
\begin{array}{ccc}
g_2 + g_3(\phi_1+\phi_2) & = & \frac{L}{y_0} \nonumber \\
g_2 + g_3(\phi_1+\phi_2) & = & \frac{L}{y_1} \nonumber
\end{array} \nonumber \\
 y_0 y_1 & = & \Lambda^4 .
\eea

We now consider the various possible solutions of these equations.

\vspace{3mm}

\noindent \underline{case 1: $L\neq0$}. This necessarily implies
that $y_0=y_1=\epsilon \Lambda^2$ with $\epsilon^2=1$. Next, we
observe that the difference between the first two equations can be
rewritten as
\be \label{u2b}
(\phi_1-\phi_2)(g_2 + g_3(\phi_1+\phi_2))=0.
\ee
Since $L\neq 0$ and $y_0$ and $y_1$ can never be zero, the second
factor in (\ref{u2b}) cannot vanish, and therefore
$\phi_1=\phi_2$. The full solution is therefore the following. Define
$\phi_0$ to be any root of the equation
\be W'(\phi_0) + 2 g_3 \epsilon \Lambda^2 =0,
\ee
then
\be \label{u2s} \phi_1=\phi_2=\phi_0,
\qquad y_0=y_1=\epsilon \Lambda^2, \qquad
L=2 \epsilon \Lambda^2 (g_2 + 2 g_3 \phi_0).
\ee
By taking the classical limit $\Lambda \rightarrow 0$, we see that
$\phi_1$ and $\phi_2$ are both the same solution of the equation
$W'(\phi)=0$. Thus, this solution describes the maximally
confining case with one solution of $W'(\phi)=0$ doubly occupied,
and the other not occupied. Classically the gauge symmetry is
still $U(2)$, but it is broken quantum mechanically to $U(1)$. There are
two different solutions depending on the choice of $\epsilon$.

Let us compare this to the expected 4d field theory answer.
According to the recipe in section~2.2, we first need to
parametrize the locus where $P(x)^2 - 4 \Lambda^4$ has a double
zero, at say $x=x_0$. With $P(x) = x^2 + s_1 x + s_2$ we need that
$P'(x_0)=0$ and $P(x_0)=-2\eta \Lambda^2$ where $\eta^2=1$. This
implies that $s_1=-2 x_0$ and $s_2=x_0^2 - 2 \eta \Lambda^2$. The
effective superpotential, expressed in terms of $s_1$ and $s_2$,
is equal to
\be W_{\rm eff} = -g_1 s_1 + \frac{g_2}{2} (s_1^2-2 s_2)
 + \frac{g_3}{3} (-s_1^3 + 3 s_1 s_2)
\ee
and when evaluated on the locus where $P(x)^2 - 4 \Lambda^4$ has a
double zero at $x=x_0$ it becomes
\be
W_{\rm eff} = 2 g_1 x_0 + g_2(x_0^2 + 2 \eta \Lambda^2) +
\frac{2 g_3}{3}(x_0^3 + 6 \eta \Lambda^2 x_0).
\ee
Extremizing this with respect to $x_0$ yields
\be
2 (W'(x_0) + 2 g_3 \eta \Lambda^2) = 0,
\ee
and for each solution $x_0$ the polynomial $P(x) = x^2 + s_1 x +
s_2$ reduces to
\be
P(x) = (x-x_0)^2 -2 \eta \Lambda^2.
\ee
This is exactly the same polynomial as the one that we obtain from
(\ref{u2p}) by substituting (\ref{u2s}), with $\eta=\epsilon$.
Thus, in the maximally confining case we reproduce the field theory answer.

\vspace{3mm}

\noindent \underline{case 2: $L=0$}. It is straightforward to find
the solution in this case. We can take for example $\phi_1$ to be
completely arbitrary, so that $\phi_2=-\phi_1-g_2/g_3$. In addition,
$y_0$ and $y_1$ are the two roots of the equation
\be \label{yeq}
y^2 +\frac{1}{g_3}W'(\phi_1) y+\Lambda^4 =0.
\ee
To understand the classical limit we send $\Lambda\rightarrow 0$
and also take $y_i\rightarrow 0$. In that limit, $\phi_1$ and
$\phi_2$ become the two distinct roots of the equation
$W'(\phi)=0$. Thus, this is the case where the gauge symmetry is
classically broken to $U(1)\times U(1)$. In contrast to the
previous situation, the quantum theory has a flat direction,
parametrized by $\phi_1$. As discussed in sections~2.4 and~3.1, 
this flat direction represents the fact
that at the quantum level there remains a $U(1)\times U(1)$ gauge
symmetry. The first, diagonal $U(1)$ is completely decoupled and
never present in the superpotential, the second $U(1)$ is
parametrized by $\phi_1$. Equivalently, we can parametrize it by
$y_0/y_1$, which is the combination that also represented this
$U(1)$ classically. The polynomial (\ref{u2p}) is equal to
\be
P(x)=x^2 +\frac{g_2}{g_3}x + \frac{g_1}{g_3}=\frac{W'(x)}{g_3} .
\ee
As expected, the $P(x)$ does not depend on the free parameter
$\phi_1$ at all. According to (\ref{pcollapse}) and (\ref{wcollapse})
we get the same result in four dimensions.
Therefore, also in this case we find complete agreement with 4d field
theory.

In summary, the structure of the vacua in our proposal
for three dimensions matches exactly the results in four
dimensions, at least for $U(2)$ with a cubic superpotential.

\subsection{$U(3)$}

Our next example is $U(3)$, again with tree level
superpotential
\be
W=g_1 \tr(\Phi) + \frac{g_2}{2} \tr(\Phi^2) + \frac{g_3}{3}
\tr(\Phi^3) .
\ee
The Lax operator is
\be \label{ll1}
M=\threea{\phi_1}{y_1}{z}{1}{\phi_2}{y_2}{y_0 z^{-1}}{1}{\phi_3} .
\ee
The polynomial $P(x)$ appearing in the spectral curve equals
\be \label{u3p} P(x) = (x-\phi_1)(x-\phi_2)(x-\phi_3) -
 (y_0 + y_1 +y_2) x +( y_2 \phi_1 + y_0 \phi_2 + y_1 \phi_3).
\ee
The resulting effective superpotential, again using (\ref{prop}),
becomes
\bea \label{ll2}
W_{\rm eff} & = &  W(\phi_1) + W(\phi_2) + W(\phi_3) +
 g_2(y_0+y_1+y_2) \nonumber \\
& & + g_3(y_0(\phi_1+\phi_3) + y_1 (\phi_1+\phi_2) + y_2 (\phi_2+\phi_3))+
  L \log\left(\frac{\Lambda^6}{y_0 y_1 y_2}\right)  .
\eea
The equations for the extrema of $W_{\rm eff}$ are

\bea \label{u3a} 
\begin{array}{ccc}
W'(\phi_1) + g_3(y_0+y_1) & = & 0 \nonumber  \\
W'(\phi_2) + g_3(y_1+y_2) & = & 0 \nonumber  \\
W'(\phi_3) + g_3(y_2+y_0) & = & 0 \nonumber  
\end{array}
& & 
\begin{array}{ccc}
g_2 + g_3(\phi_1+\phi_2) & = & \frac{L}{y_1} \nonumber \\
g_2 + g_3(\phi_2+\phi_3) & = & \frac{L}{y_2} \nonumber \\
g_2 + g_3(\phi_3+\phi_1) & = & \frac{L}{y_0} \nonumber 
\end{array} \nonumber \\
 y_0 y_1 y_2  & = & \Lambda^6 .
\eea

The first three equations can be used to solve for $y_0,y_1$ and $y_2$.
When we substitute this in the next three equations, they become
\bea
(-W'(\phi_1) + W'(\phi_2) - W'(\phi_3)) (g_2 + g_3(\phi_3+\phi_1))
& = & 2 g_3 L \nonumber \\
(-W'(\phi_1) - W'(\phi_2) + W'(\phi_3)) (g_2 + g_3(\phi_1+\phi_2))
& = & 2 g_3 L \nonumber \\
( W'(\phi_1) - W'(\phi_2) - W'(\phi_3)) (g_2 + g_3(\phi_2+\phi_3))
& = & 2 g_3 L  \label{u3h} .
\eea
An interesting simplification appears if we consider the differences
of pairs of these equations, which take the form
\be \label{u3au}
(\phi_i-\phi_j)(\phi_1+\phi_2+\phi_3-2 a_1-a_2)
(\phi_1+\phi_2+\phi_3-a_1-2a_2)=0
\ee
for some $i,j$, and where $a_1,a_2$ are the two extrema of $W'$,
\be
W'(\phi)=g_3(\phi-a_1)(\phi-a_2) .
\ee

Again, we distinguish two cases. Either all $\phi_i$ are equal, or at least
two of them are different.

\vspace{3mm}

\noindent \underline{case 1: All $\phi_i$ equal}.
This case is straightforward to analyze. All $y_i$ are identical and equal
to $\omega \Lambda^2$, with $\omega$ some third root of unity. All $\phi_i$
are also equal and a solution of the equation
\be \label{u3c}
W'(\phi_i) + 2 \omega g_3 \Lambda^3 =0 .
\ee
The polynomial in (\ref{u3p}) becomes
\be \label{u3e}
P(x)=(x-\phi_1)^3 - 3\omega\Lambda^2(x-\phi_1)
\ee
with $\phi_1$ a solution of (\ref{u3c}). For comparison with the field
theory answer, we also compute $P(x)^2 - 4\Lambda^6$, which equals
\be
P(x)^2 - 4 \Lambda^6 = (x-\phi_1-\omega^{\frac{1}{2}}\Lambda)^2
(x-\phi_1+\omega^{\frac{1}{2}}\Lambda)^2
(x-\phi_1-2\omega^{\frac{1}{2}}\Lambda)
(x-\phi_1+2\omega^{\frac{1}{2}}\Lambda) .
\ee
This has the expected form for the maximally confining case. Indeed,
in the classical limit all $\phi_i$ reduce to either $a_1$ or $a_2$,
so this situation is the one where classically the gauge group remains
unbroken. In the quantum theory the gauge symmetry is broken to $U(1)$,
and there are three different vacua depending on the choice of $\omega$.

Let us briefly check that we get the same answer in field theory.
First, we need to parametrize $s_1,s_2$ and $s_3$ in such a way
that $P(x)^2 - 4 \Lambda^6$ has two double zeroes. This
parametrization is easily found to be
\be \label{u3d}
s_1=-3 t, \qquad s_2 = 3 t^2 -3 \omega \Lambda^2, \qquad s_3 =
-t^3 + 3 \omega \Lambda^2 t .
\ee
The effective superpotential, restricted to the locus (\ref{u3d}) parametrized
by $t$ is
\be
W_{\rm eff}(t) = 3 W(t) + 3 g_2 \omega \Lambda^2 + 6 g_3 \omega \Lambda^2 t
\ee
which is extremal for $W'(t)+2 g_3 \omega \Lambda^2=0$.
With this value for $t$, we can insert (\ref{u3d}) in (\ref{u3p}) to
compute $P(x)$, and we immediately see it is the same as (\ref{u3e}).

Both for $U(2)$ as well as $U(3)$ the maximally confining case had
all $\phi_i$ equal,  and all $y_i$ equal as well. This continues to
be true for $U(N)$, as we will elaborate on in section~(\ref{qsec}).

\vspace{3mm}

\noindent \underline{case 2: some $\phi_i$ different}:
in this case (\ref{u3au}) implies that
\be \label{u3f} \phi_1+\phi_2+\phi_3 = 2 a_1 + a_2 .
\ee
We could also have chosen the left hand side to be equal to $a_1+2a_2$, but
we can then simply exchange $a_1$ and $a_2$, so there is no loss in
generality in taking (\ref{u3f}).
The single linear relation (\ref{u3f}) collapses the three equations
in (\ref{u3h}) into a single equation, since all differences vanish.
We can choose to eliminate $\phi_3$ from the equation using (\ref{u3f}).
We are then left with a single equation from (\ref{u3h}), plus the additional
equation $y_0 y_1 y_2=\Lambda^6$. Those are two equations for the three
variables $L,\phi_1,\phi_2$, and therefore there will be a free parameter
in the solution. This is similar to what we saw for $U(2)$ when it is
classically broken to $U(1)\times U(1)$. Here, as is clear from (\ref{u3f}),
we are considering the situation where $U(3)$ is classically broken to
$U(1)\times U(2)$. In the quantum theory it is further broken to $U(1)\times
U(1)$. One of these $U(1)$'s is the trivial diagonal $U(1)$, the other
is the free parameter that we are find here.

The full solution can be parametrized as follows
\bea \label{u3fs} 
\begin{array}{ccc}
\phi_1 & = & a_1 + \tilde{\phi}_1  \nonumber \\
\phi_2 & = & a_1 + \tilde{\phi}_2  \nonumber \\
\phi_3 & = & a_2 - \tilde{\phi}_1 - \tilde{\phi}_2  \nonumber
\end{array}
& & 
\begin{array}{ccc}
y_0 & = & \frac{\epsilon  \Lambda^3}{\tilde{\phi}_2 } \nonumber \\
y_1 & = & \tilde{\phi}_1 \tilde{\phi_2} \nonumber \\
y_2 & = & \frac{\epsilon  \Lambda^3}{\tilde{\phi}_1 } \nonumber
\end{array} \nonumber \\
L & = & -\epsilon g_3 \Lambda^3
\eea

with $\epsilon^2=1$ and
$\tilde{\phi}_1,\tilde{\phi}_2$ subject to one constraint
\be
\epsilon \Lambda^3 + \tilde{\phi}_1 \tilde{\phi}_2 (
 a_1-a_2 + \tilde{\phi}_1 + \tilde{\phi}_2) =0.
\ee
To compare with field theory we note that (\ref{u3fs}) substituted in
(\ref{u3p}) yields
\be
P(x)=(x-a_1)^2(x-a_2) + 2 \epsilon \Lambda^3
\ee
from which we deduce
\be
P(x)^2 - 4 \Lambda^6 = (x-a_1)^2 ( \frac{1}{g_3^2}W'(x)^2 +
 4 \epsilon \Lambda^3(x-a_2) ).
\ee
This is exactly the form that one would get from a field theory analysis
(see section~2.2), 
and therefore the result agrees with the four-dimensional field theory
expectation. Notice that the two solutions of $\epsilon^2=1$ correspond
to two inequivalent quantum vacua, as expected for $U(3)\rightarrow
U(1) \times U(2)$.

Thus, for $U(3)$ we also reproduce the complete vacuum structure that
we expect from four dimensions.

\subsection{$U(N)$ with quadratic superpotential}
\label{qsec}

The analysis for a $U(N)$ theory with a quadratic superpotential,
\begin{equation}
    W = g_1 \tr(\Phi) + \frac{g_2}{2} \tr(\Phi^2) + L \log \left ( \frac{\Lambda^{2N}}{y_0 \ldots y_{N-1}}\right )
    \label{eq:potential_phi2}
\end{equation}
can be done for general $N$. Substituting the Lax matrix for the adjoint scalar $\Phi$ into the superpotential gives the following effective superpotential
\begin{displaymath}
    W_{\mathrm{eff}} = \sum_{i=0}^{N-1} (g_2 y_i +  g_1 \phi_{i+1} + \tfrac{1}{2} g_2 \phi_{i+1}^2) + L \log \left ( \frac{\Lambda^{2N}}{y_0 \ldots y_{N-1}}\right )
\end{displaymath}
with the equations for the extrema
\begin{displaymath}
    y_0 \ldots y_{N-1} = \Lambda^{2N}
\end{displaymath}
\begin{displaymath}
    g_2 - \frac{L}{y_i} =0 \Rightarrow y_i = \frac{L}{g_2}
\end{displaymath}
\begin{displaymath}
    g_2 \phi_i + g_1=0
\end{displaymath}
From the second equation we can learn that the y's are all equal, this enables us to solve for $L$ using the first equation
\begin{equation}
    L = \epsilon g_2 \Lambda^{2},\qquad \epsilon^N=1.
    \label{eq:L_quadratic}
\end{equation}
Substituting $L$ into the equation for $y_i$ yields
\begin{displaymath}
    y_i = y = \Lambda^2 \epsilon.
\end{displaymath}
Further, we also see that all the $\phi$'s occupy the root $x=-g_1/g_2$
of $W'(x)$
\begin{displaymath}
    W'(x) = g_2 (x + \frac{g_1}{g_2}).
\end{displaymath}
The $\phi$'s should always occupy the roots of $W'(x)$ classically ($\Lambda \rightarrow 0$), but here it is even true in the quantum case.

The superpotential in the extrema is then
\begin{displaymath}
    W_{\mathrm{extr}} = N (y - \frac{g_1^2}{2 g_2})  = N (\Lambda^2 \mathrm{e}^{\frac{2\pi\mathrm{i} l}{N}} - \frac{g_1^2}{2 g_2}).
\end{displaymath}

To compare with field theory results we compute the characteristic polynomial $P_N(x) = \det (x 1_N - \Phi)$. According to appendix \ref{app:recurrence} this is either a Chebyshev polynomial of the first or of the second kind. By evaluating $P_1(x)$ and $P_2(x)$ one can see that one has to pick the polynomials of the first kind
\begin{displaymath}
    P_N(x) = 2 y^{N/2} T_N \left ( \tfrac{x+\tfrac{g_1}{g_2}}{2\sqrt{y}} \right ).
\end{displaymath}
These $P_N(x)$ are in perfect agreement with field theory results
\begin{displaymath}
\begin{split}
    P_N(x)^2-4 \Lambda^{2N} &= 4 \Lambda^{2N} \left (1-T^2_N \left (\tfrac{x+\tfrac{g_1}{g_2}}{2\sqrt{y}} \right ) \right ) = 4 \Lambda^{2N} \left ( \left (\tfrac{x+\tfrac{g_1}{g_2}}{2\sqrt{y}} \right )^2-1 \right ) U_{N-1}^2 \left (\tfrac{x+\tfrac{g_1}{g_2}}{2\sqrt{y}} \right  )\\ &= \Lambda^{2(N-1)} \epsilon^{N-1}\left ( \tfrac{1}{g_2^2}(W')^2 - 4 \Lambda^2 \epsilon \right) U_{N-1}^2 \left (\tfrac{x+\tfrac{g_1}{g_2}}{2\sqrt{y}} \right ).
\end{split}
\end{displaymath}

\subsection{Maximally confining vacua}
\label{secconf}

In the previous sections we found all vacua for some simple potentials and small rank of $U(N)$. In this section we consider the simple vacua with $\phi_i$ taken to be equal. These are considered for arbitrary $N$ and for polynomial potentials of degree $n \leq N+1$. These vacua are maximally confining.

The maximally confining vacua of a U(N) theory with superpotential 
\begin{equation}
W(x) = \sum \limits_{k=0}^{n} \frac{g_{k+1}}{k+1} x^{k+1}
\label{Wcoefs}
\end{equation} 
are obtained by taking all $y_i = (\eta \Lambda)^2$ with $\eta^{2N} = 1$, and taking all $\phi_i = \phi$, where $\phi$ is such that $\tr \; W'(M) = 0$. 
Because of the tridiagonal form of $M$ this requirement on $\phi$ is equivalent to the condition $c_0 = 0$ for the coefficient in the Laurent expansion
\begin{equation}
W(\phi + \xi + y / \xi) = \sum \limits_{k=-n}^{n} c_k \xi^k.
\label{WLaurent}
\end{equation}
Expressed in terms of the coefficients of the superpotential, the condition $c_0 = 0$ reads
\begin{equation}
\sum_{k=0}^{n} g_{k+1} \left( \sum_{l=0}^{\lfloor k/2 \rfloor} \frac{k!}{(l!)^2 (k-2l)!} \phi^{k-2l} y^l \right) = 0.
\end{equation}
with $\lfloor k \rfloor := \mathrm{max} \{ m \in \mathbb{Z} | m \leq k\}$. 
This can be expressed in terms of a hypergeometric function
\begin{equation}
\sum \limits_{k=0}^n g_{k+1} \phi^k \; _2 F_1 (-\frac{k}{2},-\frac{k}{2} + \frac{1}{2};1; 4 y/\phi^2) = 0 .
\label{critical_phi}
\end{equation}

On the other hand, according to (\ref{pcollapse}) and (\ref{wcollapse}) 
the factorization of the polynomial $P_N(x)$ appearing in the spectral curve imposes an apparently different condition on $\phi$. 
With all $\phi_i$ equal, $P_N^2(x)-4\Lambda^{2N}$ must have $N-1$ double zeroes:
\begin{equation}
P_N^2(x) - 4\Lambda^{2N} = T(x) H_{N-1}^2(x).
\end{equation} 
In this case $P_N$ is a Chebyshev polynomial: $P_N(x) = 2(\eta \Lambda)^N T_N\left(\frac{x-\phi}{2\eta \Lambda}\right)$, with $\eta^{2N} = 1$. 
Together with the convenient definition of the Chebyshev polynomials of first and second kind
\begin{equation}
 \begin{split}
 T_n(\cos x) &= \cos nx \\
 U_n(\cos x) &= \frac{\sin \; (n+1)x }{\sin x}
 \end{split}
\end{equation}
this factorization is directly observed
\begin{equation}
\frac{P_{N}^2(x) - 4\Lambda^{2N}}{4 \Lambda^{2N}} = \left( T_N^2(\frac{x-\phi}{2\eta \Lambda}) -1 \right) =  \left[ \left(\frac{x-\phi}{2\eta \Lambda}\right)^2 -1 \right] U_{N-1}^2\left(\frac{x-\phi}{2\eta \Lambda}\right).
\end{equation}
So after rescaling one obtains
\begin{equation}
T(x) = x^2 - 2 \phi x + \phi^2 - 4 y
\end{equation}
for any maximally confining vacuum, irrespective of the rank of the gauge group.

From the point of view of four dimensional gauge theory, it is expected that $T(x)$ must be related to the superpotential as
\begin{equation}
T(x) \; G_{n-1}^2(x) = (W'(x))^2 + f_{n-1}(x)
\label{swmatch}
\end{equation}
for some polynomials $G_{n-1}(x)$, $f_{n-1}(x)$ of degree $n-1$. 
Equation (\ref{swmatch}) can be seen as a set of $2n+1$ conditions that the coefficients of each power of $x$ match. 
Of these conditions $2n$ can be satisfied by appropriate choice of coefficients of $G_{n-1}(x)$ and $f_{n-1}(x)$. 
The final remaining condition turns out to be $c_0=0$, in (\ref{WLaurent}), which relates $\phi$ and $y$ to the superpotential at criticality.

To see this, examine the conditions that  $G_{n-1}(x)$ and $f_{n-1}(x)$ can be found such that (\ref{swmatch}) is true.  
As a matter of convenience, we do a field redefinition 
\begin{equation}
M \mapsto M - \phi I
\end{equation}
such that at criticality the redefined matrix $M$ has $\phi = 0$ on the diagonal. 
Next a choice of scale is made such that $y = 1$.
With these choices $T(x) = x^2 -4$. 
By a substitution 
\begin{equation}
x \mapsto \xi + \xi^{-1} 
\label{xtoxi}
\end{equation}
this can be written as a complete square
\begin{equation}
T = (\xi - \xi^{-1})^2 .
\end{equation}
So the entire left hand side of (\ref{swmatch}) is a complete square. 

Note that on the r.h.s. of (\ref{swmatch}) the coefficients of terms proportional to $x^n$ and higher powers of $x$ are entirely determined by $W'(x)$. 
The polynomial $f_{n-1}(x)$ only serves to match the coefficients of lower powers of $x$. 
Taking the square root of (\ref{swmatch}) yields
\begin{equation}
\sqrt{T(x) G_{n-1}^2(x)} = W'(x) + \frac{1}{2} \frac{f_{n-1}(x)}{W'(x)} + \ldots = W'(x) + \mathcal{O}(x^{-1})
\end{equation}
With the  substitution (\ref{xtoxi}) , noting $x^{-1} \mapsto (\xi + \xi^{-1})^{-1} = \xi^{-1}/(1+\xi^{-2}) = \mathcal{O}(\xi^{-1})$, it is observed that (\ref{swmatch}) implies
\begin{equation}
(\xi - \xi^{-1}) G_{n-1}(\xi + \xi^{-1}) = W'(\xi + \xi^{-1}) + \mathcal{O}(\xi^{-1}). 
\label{swLaurent}
\end{equation}

In order that a polynomial $G_{n-1}(x)$ of degree $n-1$ exists, the right hand side of (\ref{swLaurent}) must be divisible by $(\xi - \xi^{-1})$. 
Choosing the $\mathcal{O}(\xi^{-1})$ polynomial in $\xi^{-1}$ as follows almost guarantees this:
\begin{equation}
(\xi - \xi^{-1}) G_{n-1}(\xi + \xi^{-1}) = W'(\xi + \xi^{-1}) - 2 [ W'(\xi + \xi^{-1}) ] _{-}. 
\end{equation}
Here $[g(\xi)]_{-}$ denotes the part of the Laurent series of $g$ with strictly negative powers of $\xi$. 
The right hand side, written as a Laurent series in $\xi$ is of the form
\begin{equation}
r.h.s. = c_0 + \sum \limits_{k=1}^{n} c_{k} (\xi^{k}- \xi^{-k}).
\end{equation} 
This is divisible by $(\xi - \xi^{-1})$ iff $c_0=0$, with $c_0$ the coefficient of $\xi^{0}$ in the Laurent series of $W'(\xi + \xi^{-1})$. 
This is exactly the requirement for criticality, with the field redefinition and choice of scale that set $\phi = 0$ and $y=1$.  

In fact, a concrete expression for the coefficients of $G_{n-1}(x)$ can be found, keeping $\phi$ and $y$ explicit.  
In order that (\ref{swmatch}) be true, the coefficients of each power of $x$ must match. 
As noted earlier, by choosing $f_{n-1}(x)$ appropriately, $n$ such coefficients can be matched. 
This leaves $n+1$ coefficients to be matched, $n$ of which can be made to do so by an appropriate choice of $G_{n-1}(x)$ \footnote{note the index: this corresponds to an expansion $\sum_{i=0}^{n-1} g_{i+1} x^i$}. 
In the end there remains a single relation between $\phi$, $y$ and the coefficients $g_k$ which needs to be satisfied for (\ref{swmatch}) to be true.

Now let us check that the single non-trivial relation is precisely (\ref{critical_phi}). 
Define
\begin{equation}
 \begin{split}
  (W'(x))^2 &= \sum \limits_{k=0}^{2n} \omega_k x^k = \sum \limits_{k=0}^{2n} x^{k} \sum \limits_{0 \leq r,s \leq s \; ; \; r+s = k} g_{r+1} g_{s+1} \\
  G_{n-1}^2(x) &= \sum \limits_{k=0}^{2n-2} \gamma_k x^k = \sum \limits_{k=0}^{2n-2}x^{k}  \sum \limits_{0 \leq r,s \leq n-1 \; ; \; r+s = k} \chi_{r+1} \chi_{s+1}.
 \end{split}
\end{equation}
In terms of $\chi_i$ and $g_i$ the relevant coefficients are
\begin{equation}
 \begin{split}
  \omega_{n+k} &= \sum \limits_{l=k}^{n} g_{l+1} g_{n+k-l+1} \quad (k=0,\dots,n)\\
  \gamma_{n+k} &= \sum \limits_{l=k+1}^{n-1} =\chi_{l+1} \chi_{n+k-l+1} \quad (k=0,\dots,n-1)
 \end{split}
\end{equation}
The objective is to find coefficients $\chi_i$ such that (\ref{swmatch}) is true, as far as the coefficients of the powers $x^{n+1}$ to $x^{2n}$ are concerned. 
That is
\begin{equation}
 \begin{split}
  \omega_{2n} &= \gamma_{2n-2} \\
  \omega_{2n-1} &= \gamma_{2n-3} -2\phi \gamma_{2n-2} \\
  \omega_{n+k} &= \gamma_{n+k-2} -2 \phi \gamma_{n+k-1} + (\phi^2-4y) \gamma_{n+k} \quad (k=1,\dots,n-2) .
 \end{split}
\label{squarecoefs}
\end{equation}
The coefficients of powers $x^0$ to $x^{n-1}$ can be matched by appropriate choice of the polynomial $f_{n-1}$ in (\ref{swmatch}). 
So in the end a single nontrivial relation remains, relating $\phi$, $y$ and the coefficients of $W'$.

The $\chi_i$ can be solved one by one, starting from $\chi_{n}$. 
The top two equations in (\ref{squarecoefs}) have a solution 
\begin{equation}
 \begin{split} 
  \chi_{n} &= -g_{n+1} \\
  \chi_{n-1} &= -(g_{n} + \phi g_{n+1})
 \end{split}
\end{equation}
The next $n-3$ equations are solved by
\begin{equation}
\chi_{r+1} = - \sum \limits_{l=0}^{n-1-r} g_{r+2+l} \phi^l \; _2 F_1 (- \frac{l}{2}, - \frac{l}{2} +\frac{1}{2} ; 1 ; 4 y/\phi^2 ) .
\label{mchyper}
\end{equation}
Subsequent $\chi_r$'s can be solved one after another because a $\chi_r$ with smaller $r$ appears only in coefficients of lower powers of $x$. 
Thus every next condition, on the coefficients of ever lower powers of $x$ is solved by appropriate choice of $\chi_r$ with ever lower indices. 
At some point, this process stops, as there is no $\chi_{0}$; the coefficient of $x^{-1}$ in $G_{n-1}(x)$ vanishes. 
So in order that the coefficients of $x^n$ in (\ref{swmatch}) match, (\ref{mchyper}) must be satisfied with $\chi_{0}=0$.
Note that this relation is precisely (\ref{critical_phi}) relating $\phi$ to $y$ and the coefficients of the superpotential.

\subsection{Lifting solutions from U(N) to U(tN)}
\label{secspec}

It is known \cite{civ} that supersymmetric vacua of a $U(N)$ gauge theory
with superpotential $W$ can be lifted to supersymmetric vacua of a $U(tN)$ gauge
theory with exactly the same superpotential. On the level of the Seiberg-Witten
polynomial, this lifting involves Chebyshev polynomials and a few other 
ingredients. It turns out, as we will show in this section, that it is very simple
to do this at the level of Lax operators. One simply takes $t$ copies of the
Lax operator of $U(N)$ to construct a Lax operator of $U(tN)$ that is periodic
in steps of $N$. This new Lax matrix is an extremum for the same superpotential,
and this replica trick therefore provides a very simple picture of how to lift
vacua. To show technically how this works, we start with 
the Lax operator for U(N) 
\begin{equation}
M_{N} =
 \begin{pmatrix}
  \phi_1        &   y_1   &   0    & \dots & z         \\
   1            & \phi_2  &  y_2   & \dots & 0         \\
   0            &    1    & \phi_3 & \dots & 0         \\
   .            &    .    &  .     & \dots & y_{N-1}   \\
   y_{0} z^{-1} &    .    &  .     &   1   & \phi_N
 \end{pmatrix}
\label{UNmatrix}
\end{equation}
and the corresponding polynomial $P_{N}(x)$ in the spectral curve 
\begin{equation}
P_{N}(x) = \det(x I_{N} - {M_N}) + z + \Lambda^{2N} z^{-1},
\label{UNpolynomial}
\end{equation}
where the energy scale $\Lambda^2 \geq 0$ sets the condition
\begin{equation}
\Lambda^{2N} = \prod \limits_{i=0}^{N-1} y_i .
\label{LagrangeN}
\end{equation}

As explained above, for a U(tN) theory with a superpotential $W$ of degree $d \leq N+1$ a special form of $M_{tN}$ can be considered so that the analysis can be reduced to that of the U(N) case, with the same superpotential.
This is possible when the entries of $M_{tN}$ are periodically identified like
\begin{equation}
 \begin{split}
  \phi_{i} & \equiv \phi_{i+N} \\
  y_{i}    & \equiv y_{i+N}
 \end{split}
\end{equation}
so that the only non zero entries of $M_{tN}$ are
\begin{equation}
 \begin{split}
  (M)_{i,i}   &= \phi_{i} \\
  (M)_{i,i-1} &= 1 \\
  (M)_{i,i+1} &= y_{i} \\
  (M)_{1,tN}  &= z \\
  (M)_{tN,1}  &= y_{0} z^{-1}
 \end{split}
\end{equation}
The condition set by the energy scale is written as
\begin{equation}
\prod \limits_{i=0}^{tN-1} y_i = \Lambda^{2tN} \geq 0.
\label{LagrangetN}
\end{equation}
Note in particular that the (physical) energy scale of the $U(tN)$ theory should be real. This condition can be satisfied by a family of $t$ different complex valued $\Lambda^{N}$ in the U(N) theory,  $\{\Lambda^{N}, \eta^2 \Lambda^{N}, \dots, \eta^{2(t-1)} \Lambda^{N} \}$, with $\eta$ a $2t$-th root of unity.

The polynomial $P_{tN}(x)$ in the spectral curve of $M_{tN}$ is
\begin{equation}
P_{tN}(x) = \det( xI_{tN} - M_{tN}) + z + \Lambda^{2tN} z^{-1}
\end{equation}
The determinant can be calculated by considering a gauge equivalent matrix.
By a gauge transformation $M_{tN}$ can be brought into a form that is invariant under cyclic permutations of order N.
We define
\begin{equation}
G = \mathrm{diag} (1, z^{1/tN} , z^{2/tN} , \dots , z^{(tN-1)/tN})
\end{equation}
then $\tilde{M}_{tN} = G M_{tN} G^{-1}$ has factors of $z^{1/t}$ and $z^{-1/t}$ democratically distributed over the $(\tilde{M})_{i,i-1}$ and  $(\tilde{M})_{i,i+1}$ entries respectively.
$\tilde{M}_{tN}$ satisfies
\begin{equation}
S \tilde{M}_{tN} S^{-1} =  \tilde{M}_{tN}
\label{SMS=M}
\end{equation}
where
$$
S =
 \begin{pmatrix}
   0     && I_{(t-1)N} \\
   I_{N} && 0
  \end{pmatrix}
$$
is the matrix that generates a cyclic permutation of order N on tN elements,
$S^{-1} = S^{T}$.

In case the degree of the superpotential is small enough, $\mathrm{deg}(W) \leq N+1$, the cyclic invariance, (\ref{SMS=M}), ensures that the equations of motion collapse to those of the U(N) theory with the same superpotential.
First, because in $\tilde{M}_{tN}$ the $\phi_{i}$ and $y_{i}$ appear only linearly, derivatives with respect to $\phi_{i+kN}$ and $y_{i+kN}$ can be replaced with derivatives with respect to $\phi_i$ and $y_i$
\begin{equation}
 \begin{split}
  \partial_{\phi_{i+N}} \mathrm{tr} \left( \tilde{M}_{tN}^{n+1} \right)
  &= (n+1) \mathrm{tr} \left( \tilde{M}_{tN}^{n} \partial_{\phi_{i+N}}  \tilde{M}_{tN} \right) \\
  &= (n+1) \mathrm{tr} \left( S \tilde{M}_{tN}^{n}  S^{-1}  S ( \partial_{\phi_{i+N}}    \tilde{M}_{tN}) S^{-1} \right) \\
  &= (n+1)\mathrm{tr} \left( \tilde{M}_{tN}^{n}  ( \partial_{\phi_{i}} \tilde{M}_{tN}) \right) \\
  &= \partial_{\phi_{i}} \mathrm{tr} \left( \tilde{M}_{tN}^{n+1} \right) .
 \end{split}
\end{equation}
Second, in $\tilde{M}_{tN}$ the $\phi_i$ and $y_i$ appear only on the diagonal and the superdiagonal respectively
\begin{equation}
(\tilde{M}_{tN})_{ij} = \phi_i \delta_{i,j} + y_{i} z^{-1/tN} \delta_{i, j-1} +z^{1/tN} \delta_{i,j+1}
\end{equation}
the diagonal elements of $\tilde{M}_{tN}^{n}$ each depend on at most $n$ consecutive $y_i$'s and $\phi_i$'s.
Therefore, all equations of motion for $\phi_{1}$ up to $\phi_{tN}$ can be both mapped to equations of motion for $\phi_{1}$ to $\phi_{N}$ and these also consistently collapse onto the equations for the first $N$ $\phi_i$'s.
The same holds for the equations of motion for all $y_i$'s.
Also the equation of motion for the Lagrange multiplier, (\ref{LagrangetN}) maps to (\ref{LagrangeN}).

$\tilde{M}_{tN}$ can be explicitly written as a $t \times t$ matrix of which each entry is itself one of four $N \times N$ matrices
\begin{equation}
\tilde{M}_{tN} =
 \begin{pmatrix}
 A     & D     & 0     & \dots & \dots & 0     & E     \\
 E     & A     & D     & 0     & \dots & \dots & 0     \\
 0     & E     & A     & D     & \dots & \dots & \dots \\
 \dots & 0     & E     & A     & \dots &  \dots & \dots \\
 \dots & \dots & \dots & \dots & \dots & \dots & \dots \\
 \dots & \dots & \dots & \dots & \dots & \dots & D     \\
  D    & 0     & \dots & 0     & \dots & E     & A
 \end{pmatrix}.
\end{equation}
The $N \times N$ matrices A, D, and E are of the following form
\begin{equation}
A =
 \begin{pmatrix}
  \phi_1   & y_1 z^{-1/tN} & 0            & \dots &  \dots & 0     \\
   z^{1/tN} & \phi_2       & y_2 z^{-1/tN} & 0     & \dots  & 0     \\
   0       &     z^{1/tN}  & \dots        & \dots &  \dots & \dots \\
   \dots   &  0           &  \dots       & \dots & \dots  & \dots  \\
   \dots   & \dots        & \dots        & \dots & \dots  & \dots \\
   \dots   & \dots        & \dots        & \dots & \dots  &y_{N-1} z^{-1/tN}  \\
    0      & \dots        & \dots        & \dots & z^{1/tN}& \phi_{N} \\
 \end{pmatrix}
\end{equation}

\begin{equation}
D = y_{0} z^{-1/tN}
   \begin{pmatrix}
   0_{1 \times (N-1)} & 0_{(N-1) \times (N-1)} \\
   1                  & 0_{(N-1) \times 1}
   \end{pmatrix}
\end{equation}

\begin{equation}
E = z^{1/tN}
   \begin{pmatrix}
   0_{1 \times (N-1)}        & 1 \\
   0_{(N-1) \times (N-1)}    & 0_{(N-1) \times 1}
   \end{pmatrix}
\end{equation}
Because of (\ref{SMS=M}), there exists a basis of simultaneous eigenvectors of $\tilde{M}_{tN}$ and $S$.
The eigenvectors of  $S$ fall into $t$ N-dimensional subspaces, each of which is labeled by a different $t$-th root of unity $\omega^{r}$, $\omega^t=1$.
The eigenvalue equation for $\tilde{M}_{tN}$ is written in a basis of  $S$ eigenvectors $v^{(\alpha)}_r = (v^{(\alpha)}, \omega^r v^{(\alpha)} , \omega^{2r} v^{(\alpha)}, \dots , \omega^{(t-1)r} v^{(\alpha)})$, with $\alpha=1,2,\dots, N$.

For each $r$ the eigenvalue equation $\tilde{M}_{tN} v^{(\alpha)}_r = \lambda^{(\alpha)}_r v^{(\alpha)}_r$ becomes
\begin{equation}
(A + \omega^r D + \omega^{-r} E )  v^{(\alpha)}_r = \lambda^{(\alpha)}_r v^{(\alpha)}_r
\end{equation}
and therefore
\begin{equation}
\det(M_{tN}) = \det(\tilde{M}_{tN}) = \prod_{r=1}^{t} \det \left( A +  \omega^r D + \omega^{-r} E \right)
\end{equation}
The U(tN) polynomial reads
\begin{equation}
 \begin{split}
P_{tN}(x)
 &= \prod_{r=1}^{t} \det \left( xI_{N} - A -  \omega^r D - \omega^{-r} E \right) + z + \Lambda^{2tN} z^{-1} \\
 &= \prod_{r=1}^{t}  \left( P_{n}(x) - \omega^{-r} z^{1/t}  - \eta^2 \Lambda^{2N} \omega^r z^{-1/t} \right) + z + \Lambda^{2tN} z^{-1}
 \end{split}
\label{PtNPN}
\end{equation}

The polynomials $P(x)$ are by construction independent of $z$.
A convenient choice to evaluate (\ref{PtNPN}) is to take $z^{1/t} = e^{\frac{i \pi}{t}} | \Lambda^{\frac{1}{2tN}}|$.
Recall that $\omega$ is a $t$-th root of unity such that $\omega^r$ hits all $t$ different eigenvalues of the $\mathbb{Z}_t$ cyclic permutation matrix $S$.
Hence $P_{tN}(x)$ can be written as
\begin{equation}
 \begin{split}
P_{tN}(x) &=\prod \limits_{r=1}^{t} \left( P_{N}(x) - \eta \Lambda^{N} \left( e^{i \pi \frac{4r-1}{2t}} + e^{-i \pi \frac{4r-1}{2t}}   \right) \right) \\
          &= 2^t \eta^t \Lambda^{tN} \prod \limits_{r=1}^{t} \left( \frac{1}{2 \eta \Lambda^{N} } P_{N}(x) - \cos(\frac{4r-1}{2t}\pi) \right).
 \end{split}
\end{equation}

The latter expression defines the $n$-th Chebyshev polynomial of the first kind, which is defined as
\begin{equation}
T_{t}(x) = 2^{n-1} \prod \limits_{k=1}^{t} \left(x - \cos(\frac{2k-1}{2t}\pi) \right)
\label{Cheby1_n}
\end{equation}
with
\begin{equation}
s := \mathrm{max} \{ \sigma \in \mathbb{Z} : 2 \sigma \leq t \}
\end{equation}
equation (\ref{Cheby1_n}) can also be written as
\begin{equation}
 \begin{split}
 T_{n}(x) &= 2^{n-1}\prod \limits_{p=1}^{s} \left( x - \cos(\frac{2(2p)-1}{2t}\pi) \right)\prod \limits_{q=1}^{s} \left( x - \cos(\frac{2(2q-1)-1}{2t}\pi) \right) \\
 &=  2^{n-1}\prod \limits_{p=1}^{s} \left( x - \cos(\frac{4p-1}{2t}\pi) \right)\prod \limits_{q=1}^{s} \left( x - \cos(\frac{4(t-q)+3}{2t}\pi) \right) \\
 &=  2^{n-1}\prod \limits_{r=1}^{t} \left( x - \cos(\frac{4r-1}{2t}) \right)
  \end{split}
\end{equation}
so
\begin{equation} \label{replicaP}
P_{tN} = 2 \Lambda^{tN} \eta^t T_{t} \left(\frac{P_{N}(x)}{2 \eta \Lambda^N} \right)
\end{equation}
Thus the periodic ansatz for the $U(tN)$ theory yields $t$ times the number of vacua found in the $U(N)$ theory.
The polynomial $P_{tN}$ in (\ref{replicaP}) agrees precisely with what had been
found in field theory in \cite{civ}.

\subsection{U(4)}

We are now in a position to use the techniques of the preceding section to study our final example $U(4)$, again with a cubic superpotential
\be
W=g_1 \tr(\Phi) + \frac{g_2}{2} \tr(\Phi^2) + \frac{g_3}{3}
\tr(\Phi^3) .
\ee
The Lax operator, polynomial $P(x)$, superpotential $W_{\rm eff}$
and equations of motion are straightforward generalizations of
(\ref{ll1}), (\ref{u3p}), (\ref{ll2}) and (\ref{u3a}), with
$\Lambda^6$ replaced by $\Lambda^8$. Instead of giving a lengthy
and tedious analysis of the possible solutions of the equations of
motion, we will simply present a solution for each of the critical
points that we expect, based on the knowledge of the possible
solutions in four dimensions. These are given explicitly in
section~3.3 of \cite{csw}.

To describe the qualitative form of the solutions, we write
$W'(x)=(x-a_1)(x-a_2)$. Classically, each of the $\phi_i$ is equal
to either $a_1$ or $a_2$. We denote the number of $\phi_i$ for
which $\phi_i=a_j$ by $N_j$, so that $N_1+N_2=4$. However, this
does not fully specify the different solutions. For $U(4)$ with a
cubic superpotential, we know that at a minimum $P(x)^2 - 4
\Lambda^8$ has two double zeroes. These double zeroes are
distributed over $P(x)-2\Lambda^4$ and $P(x)+2\Lambda^4$, since
these two factors cannot have a common zero. We denote by
$r_{\pm}$ the number of double zeroes in $P(x)\pm 2\Lambda^4$, so
that $r_+ + r_-=2$.

When all $\phi_i$ are equal, with $(N_1,N_2)=(4,0)$ or $(0,4)$, we
have a maximally confining vacuum, and these were already
described in detail in (\ref{secconf}).

When $(N_1,N_2)=(2,2)$, we can use the results in (\ref{secspec})
to find solutions by lifting solutions in $U(2)$ to $U(4)$. One
easily finds that these solutions have $(r_+,r_-)=(2,0)$ or
$(0,2)$. In addition, they necessarily have
$(\phi_1,\phi_2,\phi_3,\phi_4)=(a_1,a_2,a_1,a_2)$ or
$(a_2,a_1,a_2,a_1)$.

However, this does not exhaust all solutions with
$(N_1,N_2)=(2,2)$. There are also solutions that have
$(r_+,r_-)=(1,1)$. In addition, we have not yet considered
solutions with $(N_1,N_2)=(1,3)$ or $(3,1)$. It turns out that all
missing solutions are part of one family, that can be described as
follows. First, by an overall rescaling and by shifting $x$ by a
constant we can always choose $W(x)$ so that
\be W'(x)=x^2+ x \frac{\Lambda^4}{a^3} - a^2,
\ee
for some parameter $a$. The solution is then
\bea \label{u4s} 
\begin{array}{ccc}
\phi_1 & = & \frac{a^2}{\phi_3} \nonumber \\
\phi_2 & = & \frac{a^2}{\phi_4} \nonumber \\
\phi_3 & = & \phi_3 \nonumber \\
\phi_4 & = & \phi_4 \nonumber
\end{array}
& & 
\begin{array}{ccc}
y_0 & = & a^2 \frac{\phi_3 + \phi_4}{\phi_3} \nonumber \\
y_1 & = & a^2 +\frac{a^4}{\phi_3 \phi_4} \nonumber \\
y_2 & = & a^2 \frac{\phi_3 + \phi_4}{\phi_4} \nonumber \\
y_3 & = & a^2 +\phi_3 \phi_4 \nonumber 
\end{array} \nonumber \\
L & = & \frac{\Lambda^4}{a}
\eea
with $\phi_3,\phi_4$ subject to the constraint
\be
(\phi_3+\phi_4)(a^5 + a^3 \phi_3 \phi_4) + \phi_3 \phi_4 \Lambda^4
=0 .
\ee
As explained in \cite{csw}, there are two
classical limits, one is $\Lambda\rightarrow 0$ while keeping $a$
fixed, the other is $\Lambda\rightarrow 0$ while keeping
$\Lambda^4/a^3$ fixed. The first corresponds to $(N_1,N_2)=(2,2)$
and $(r_+,r_-)=(1,1)$, the second to $(N_1,N_2)=(1,3)$ or $(3,1)$.

We have accounted for all vacua that we expect in four
dimensions. One interesting feature of the solutions is that in
the case with $(N_1,N_2)=(2,2)$, all classical limits have
$\phi_1=\phi_3$ and $\phi_2=\phi_4$, but there is no solution
whose classical limit obeys $\phi_1=\phi_2$ and $\phi_3=\phi_4$.
It therefore appears that one should be careful in choosing the
right ordering of the eigenvalues, not all orderings will give
rise to a solution of the quantum equations of motion. This is not
a contradiction, since the choice of Lax matrix breaks the $S_N$
symmetry to $Z_N$, and there is no symmetry that arbitrarily
permutes the eigenvalues.

The solution in (\ref{u4s}) has one free parameter, which
corresponds to the extra $U(1)$ that appears when breaking
$U(4)\rightarrow U(2)\times U(2)$.

\section{The semi-classical expansion}

In the previous section we presented several examples of superpotentials
and their extrema. The solutions depend in a non-trivial way on $\Lambda$,
and it is important to understand the nature of the semi-classical expansion,
certainly if we want to compare our results to direct field theory 
calculations in three dimensions. A precise understanding of the
semi-classical expansion is probably also important in order to understand
the relation to the 4d description in terms of gluino bilinear superfields,
as we will discuss in section~7.

What is the general structure of the semiclassical expansion that one
would expect to find? In four dimensions in a situation where the
the gauge group is classically broken to $\prod_i U(N_i)$ the
effective superpotential explicitly takes the form of a semiclassical
expansion. In addition to the Veneziano-Yankielowicz superpotentials
(\ref{wvy}) for each unbroken $U(N_i)$, there are many additional terms
coming from the planar diagrams of the matrix model. These are polynomial and
give ultimately rise to an expansion in positive but possibly fractional
powers of $\Lambda$. The low-energy scales $\Lambda_i$ of the unbroken
$U(N_i)$ that appear in
the Veneziano-Yankielowicz superpotential (\ref{wvy}) are obtained by
scale matching and given by
\be
\label{match}
\Lambda_i^{3N_i} = g_n^{N_i} \Lambda^{2N} \prod_{j\neq i} (a_j-a_i)^{N_i-2N_j},
\ee
with $a_i$ and $g_n$ defined in (\ref{wpri}). 

On $R^3\times S^1$, the Veneziano-Yankielowicz superpotential is no longer
appropriate, the relevant superpotential is instead given by (\ref{w3})
or equivalently (\ref{w3a}). Therefore, we expect that the superpotential
on $R^3 \times S^1$ should admit an expansion of the form
\be \label{semex}
W = \sum_i \sum_{j=0}^{N_i-1} y^{(i)}_j + \mbox{\rm higher order terms}
\ee
where the $y^{(i)}_j$ are analogues of the variables $y_i$ for each of
the classically unbroken $U(N_i)$ groups, and they should therefore
obey
\be \label{pcons} \prod_{j=0}^{N_i-1} y^{(i)}_j = \Lambda_i^{3N_i}
\ee
for each $i$. The superpotential in the form (\ref{semex}) only depends
on variables $y_i$, not on the eigenvalues $\phi_i$ of $\Phi$. In addition,
it depends on a choice of classical vacuum, whereas the superpotential
(\ref{prop}) with the $\phi_i$ included describes all vacua.
Therefore, a description like (\ref{match}) can only emerge after we
integrate out the $\phi_i$. 

To illustrate how an expansion like (\ref{semex}) can emerge, we consider
the case $U(4)\rightarrow U(2)\times U(2)$, with superpotential $W(x) = 
x^3/3-a^2 x$, so that $W'(x)=(x-a)(x+a)$. 
As a first step, we integrate out the 
$\phi$'s in the quantum superpotential. This can be done explicitly,
the solutions for the $\phi_i$ read
\begin{displaymath}
        \phi_1 = \sqrt{a^2 - y_0 -y_1} \qquad \phi_3 = \sqrt{a^2-y_2-y_3}
\end{displaymath}
\begin{displaymath}
        \phi_2 = -\sqrt{a^2 - y_1-y_2} \qquad \phi_4 = -\sqrt{a^2 -y_0-y_3}.
\end{displaymath}
Notice that we chose the signs of the square roots in such a way that
in the classical limit $y_i\rightarrow 0$ we indeed end up in a vacuum where
$U(4)$ is broken to $U(2)\times U(2)$. Therefore integrating out the
$\phi_i$ also involves the selection of a classical vacuum configuration. 
If we insert the values for $\phi_i$ in the superpotential, and expand
the result to second order in $y$, we get
\begin{displaymath}
        W_\mathrm{eff} = L\log(\frac{\Lambda_4^8}{y_0 y_1 y_2 y_3}) 
- \frac{1}{2a} (y_0-y_2)(y_1-y_3)+\ldots
\end{displaymath}
If we next integrate out $L$, this becomes
\be \label{weff2}
W_\mathrm{eff} = -\frac{1}{2a} \left(
y_0 y_1 + \frac{\Lambda^8}{y_0y_1} - y_1 y_2 - \frac{\Lambda^8}{y_1 y_2}
\right).
\ee
Interestingly, this depends on only two independent variables, namely
$y_0 y_1$ and $y_1 y_2$, and therefore it is already of the form (\ref{semex}).
Indeed, if we define
\be
y^{(1)}_0 = -\frac{y_0y_1}{2a}, \qquad
y^{(1)}_1 = -\frac{\Lambda^8}{2ay_0y_1}, \qquad
y^{(2)}_0 = \frac{y_1y_2}{2a}, \qquad
y^{(2)}_1 = \frac{\Lambda^8}{2ay_1y_2}
\ee
then (\ref{weff2}) is of the form (\ref{semex}), and $y^{(1)}_0 y^{(1)}_1=
\Lambda_1^6=\Lambda^8/(2a)^2$, $
y^{(2)}_0 y^{(2)}_1= \Lambda_2^6 = \Lambda^8/(2a)^2$, completely in
agreement with (\ref{pcons}) and (\ref{match}).

Of course, the above result is simply the semiclassical result at leading order.
It would be very interesting to go beyond the leading order, and to understand
in detail how the expansion is organized. We have not studied this in detail,
but expect the following. In general the value of the superpotential is
invariant under the flows (\ref{commflow}) of the integrable system. Some
of the flows are stationary at the extremum, but some are not, and that is
why there is a $k-1$ complex parameter family of minima (see also the discussion
in section~3.1.) If we could redefine our $y$ variables in such a way that
$N-k$ of them are independent of the $k-1$ flows that do not degenerate,
then the superpotential should be a non-trivial function of these $N-k$ variables
only. This is exactly the number of independent variables that appears in the
semiclassical expansion (\ref{semex}). Hence we expect that the semiclassical
expansion appears naturally by integrating out the variables $\phi_i$ and $L$, and by
subsequently redefining the complex variables $y_i$ in a suitable way, exactly
as in our example above. 

To conclude this section, we illustrate in the case of $U(4)$ with a cubic potential
$W(x)=x^3/3-a^2 x$ how the solutions found in section~4.6 can be expanded in
(possibly fractional) powers of $\Lambda$.

For the maximally confining case, with $(N_1,N_2)=(4,0)$, we have
\bea
\phi_i & = & a - \omega \frac{\Lambda^2}{a} +\ldots \nonumber \\
y_i & = & \omega \Lambda^2 + \ldots \nonumber \\
L & = & 2 a \omega \Lambda^2 + \ldots
\eea
with $\omega$ a fourth root of unity.

For $(N_1,N_2)=(2,2)$ with $(r_+,r_-)=(2,0)$ we find
\bea
\phi_1=\phi_3=-\phi_2=-\phi_4  & = & a - (\xi +\frac{1}{\xi})
\frac{\Lambda^2}{2a}
+ \ldots \nonumber \\
y_0=y_2 & = & \xi \Lambda^2+\ldots \nonumber \\
y_1=y_3 & = & \frac{1}{\xi} \Lambda^2 +\ldots \nonumber \\
L & = & 0 \label{expans2}.
\eea
Here, and in the solutions below, $\xi$ indicates the free
parameter that is related to the additional unbroken $U(1)$ that
one gets in the corresponding vacuum solution.

For $(N_1,N_2)=(2,2)$ with $(r_+,r_-)=(1,1)$ the expansion reads
\bea 
\begin{array}{ccc}
\phi_1 & = & a - (\xi -\frac{1}{\xi}) \frac{\Lambda^2}{2a}
+ \ldots \nonumber \\
\phi_2 & = & -a - (\xi +\frac{1}{\xi}) \frac{\Lambda^2}{2a}
+ \ldots \nonumber \\
\phi_3 & = & a + (\xi -\frac{1}{\xi}) \frac{\Lambda^2}{2a}
+ \ldots \nonumber \\
\phi_4 & = & -a + (\xi +\frac{1}{\xi}) \frac{\Lambda^2}{2a}
+ \ldots \nonumber
\end{array}
& & 
\begin{array}{ccc}
y_0 & = & \xi \Lambda^2+\ldots \nonumber \\
y_1 & = & -\frac{1}{\xi} \Lambda^2 +\ldots \nonumber \\
y_2 & = & -\xi \Lambda^2+\ldots \nonumber \\
y_3 & = & \frac{1}{\xi} \Lambda^2 +\ldots \nonumber 
\end{array} \nonumber \\
L & = & \frac{\Lambda^4}{a} + \ldots .
\eea

This result differs considerably from (\ref{expans2}),
illustrating the fact that these are two different solutions.

Finally, for $(N_1,N_2)=(3,1)$ the result is an expansion in
\be
\epsilon=\left(\frac{\Lambda^4}{2a}\right)^{1/3},
\ee
which reads

\bea 
\begin{array}{ccc}
\phi_1 & = & a -\frac{\epsilon}{\xi} + \ldots \nonumber \\
\phi_2 & = & a -\frac{\epsilon^2}{a} + \ldots \nonumber \\
\phi_3 & = & a -{\epsilon}{\xi} + \ldots \nonumber \\
\phi_4 & = & -a +{\epsilon}({\xi}+\frac{1}{\xi}) + \ldots \nonumber 
\end{array}
& & 
\begin{array}{ccc}
y_0 & = & \frac{2 a \epsilon}{\xi} +\ldots \nonumber \\
y_1 & = & \epsilon^2 +\ldots \nonumber \\
y_2 & = & \epsilon^2 +\ldots \nonumber \\
y_3 & = & {2 a \epsilon}{\xi} +\ldots \nonumber 
\end{array} \nonumber \\
L & = & 2 a \epsilon^2 + \ldots \, \, .
\eea


\section{Integrable systems interpretation}

The Lax matrix of the periodic Toda chain played an essential role in the
construction of the effective superpotential in three dimensions. One of the 
motivations of this work was to try to find a direct relation between the 
periodic Toda chain and the Dijkgraaf-Vafa matrix model. We have not yet
succeeded in finding a direct relation, but nevertheless we have found that
the supersymmetric vacua in three dimensions have a very nice interpretation in
terms of the Toda integrable system. To explain this, we first need to define
the operator ${\cal L}$ that appears in (\ref{commflow}),which acts
on powers of the Lax matrix $M$; $M$,
given in (\ref{deflax}), depends on a spectral parameter $z$. Therefore, 
any power of $M$ can be written as a series in $z$ as
\be M^k = \sum_r z^r M^k_{(r)} \ee
Next, we define $M^k_+$ as the sum of the upper diagonal part of $M^k_{(0)}$ plus
$\sum_{r<0} z^r M^k_{(r)}$, $M^k_-$ as the sum of the lower diagonal part of
$M^k_{(0)}$ plus $\sum_{r>0} z^r M^k_{(r)}$, and finally $M^k_0$ as the diagonal
part of $M^k_{(0)}$. Then
\be \label{defll}
{\cal L}(M^k) \equiv M^k_- + M^k_0 - M^k_+ .
\ee
One can interpret powers of $M$ also as elements of the affine Lie algebra
$\widehat{gl}_N$, with $z^{-1}$ corresponding to the extra affine root of the
extended Dynkin diagram. Then (\ref{defll}) is nothing but the statement that 
all positive affine roots change sign. Alternatively, one can embed 
$\widehat{gl}_N$ in $gl_{\infty}$, by extending every matrix to a periodic infinite
matrix with period $N$ (i.e. $A_{i+N,j+N}=A_{ij}$), and by replacing $z$ by the
shift matrix $D^N$ with $D_{ij}=\delta_{i-1,j}$. Then (\ref{defll}) amounts
to just changing the sign of the upper triangular part of the corresponding infinite
matrix.

The equations of motion derived from (\ref{prop}) imply that the flow 
generated by $W(M)$ degenerates. This translates to the equations
\be \label{weq1}
[M,W'(M)_+ - W'(M)_-]=0, \qquad W'(M)_0=0.
\ee
A second observation, which follows from (\ref{defspec}), is
that 
\be \label{weq2} 
[M,P_N(M)_+ - P_N(M)_-]=0.
\ee

The two equations (\ref{weq1}), (\ref{weq2}) define a reduction of the integrable
Toda system. As we will show in \cite{toappear}, these equations imply the 
existence of polynomials $H_{N-k},T_{2k},G_{n-k},f_{n-1},U_k$ of degrees $N-k,2k,n-k,n-1,k$ for $0 < k \leq n$ such that
\bea
(W'(M)_+ - W'(M)_-)^2 & = & W'(M)^2 + f_{n-1}(M)  \label{c1} \\
W'(M)_+ - W'(M)_- & = & G_{n-k}(M) (U_k(M)_+ - U_k(M)_-) \label{c2} \\
(P_N(M)_+ - P_N(M)_-)^2 & = & P_N(M)^2 - 4 \Lambda^{2N} \label{c3} \\
P_N(M)_+ - P_N(M)_- & = & H_{N-k}(M) ((U_k(M)_+ - U_k(M)_-) \label{c4} \\
(U_k(M)_+ - U_k(M)_-)^2 & = & T_{2k}(M). \label{c5}
\eea

This system of equations demonstrates not only that the extrema of the superpotential
(\ref{prop}) are in exact one-to-one correspondence with the four-dimensional results,
it also shows that we can construct the four-dimensional equations (\ref{pcollapse})
and (\ref{wcollapse}) directly in terms of the Lax matrix. In particular, the matrix
model resolvent \cite{dv} $2R(z)=-\sqrt{W'(z)^2 + f_{n-1}(z)} + W'(z)$ satisfies the
very simple equation
\be \label{mres} R(M) =W'(M)_-.
\ee
Actually, equation (\ref{weq1}) is tantalizingly close to a similar equation that
can be derived for matrix integrals of the form 
\be \label{mint} \int d\Phi e^{-W(\Phi)}.
\ee
If one defines a set of orthogonal polynomials with respect to the measure
$e^{-W(x)}$, of the form $p_n=x^n+\ldots$, one can construct a semi-infinite
matrix $Q$ that acts on $p_n$ as multiplication by $x$. This semi-infinite 
matrix is tridiagonal (i.e. $Q_{ij}=0$ for $|i-j|>1$), and obeys the ``string
equation''
(see for example \cite{review} and references therein)
\be \label{qeq}
[Q,-\frac{1}{2}(W'(Q)_+ - W'(Q)_-)]=1 .
\ee
The only difference between (\ref{weq1}) and (\ref{qeq}) is that one is replaced by 
zero. Equation (\ref{qeq}) is also the equation that gives rise to the string equation
in the matrix model description of minimal models coupled to gravity. 
Though (\ref{weq1}) and (\ref{qeq}) are very similar, we have not found a direct
map between $Q$ and $M$. Whereas $Q$ is relevant for the orthogonal polynomials
defined with respect to $e^{-W(x)}$, $M$ seems to define orthogonal polynomials for 
a measure which coincides with the gauge theory resolvent instead. 
Various other relations between matrix integrals and the Toda lattice equations are
discussed in e.g. \cite{mathco0010135,mathph0209019}.

The definition of the gauge theory resolvent (the generating functional of
$\langle {\rm Tr}(\Phi^m) \rangle$) requires some discussion. For sufficiently large $m$,
we can no longer have the identity $\langle \Phi^m \rangle={\rm Tr}(M^m)$, because
the right hand side will start to depend non-trivially on the spectral parameter.
The resolution is to use the fact that given a Lax matrix $M_N$ of size
$N\times N$, there is a corresponding Lax matrix $M_{tN}$ of size $tN\times tN$
which is constructed using the replica procedure given in section~4.5. For
small $m$, ${\rm Tr}(M_N^m)=t^{-1} {\rm Tr}(M_{tN}^m)$, but for larger values
of $m$ the right hand side starts to depend on the spectral parameter, whereas
the left hand side does not. We therefore propose
\be \label{defres}
\langle \Phi^m \rangle = \lim_{t\rightarrow \infty} \frac{1}{t}
{\rm Tr}(M_{tN}^m) 
\ee
as the right definition for arbitrary $m$. 
Note that the replica procedure yields 
\begin{equation}
P_{tN}(x)=2 \Lambda^{tN}T_{t}(\frac{P_{N}(x)}{2 \Lambda^{N}}).
\end{equation}
With this expression for $P_{tN}(x)$ and the identities for Chebyshev polynomials 
\begin{equation}
 \begin{split}
  T_{n}'(x)&=n U_{n-1}(x)\\
  T^2_{n}(x)-1 &= (x^2-1) U^2_{n-1}(x)
 \end{split}
\end{equation}
it is easy to see that this proposition yields a resolvent that agrees with the field theory result of \cite{cdsw}, 
\begin{equation}
\langle \mathrm{tr}\frac{1}{x-\Phi}\rangle = \frac{P'(x)}{\sqrt{P^2(x)-4\Lambda^{2N}}}.
\end{equation}
 Therefore, $M_{tN}$ for large
$t$ provides a master field for $\Phi$. For the simple case of a quadratic
superpotential (see section~4.3) $M$ takes a simple form, and $M_{tN}$ for
infinite $t$ can be written in a simple way in terms of Cuntz variables. In this way it
also provides a master field for the Gaussian matrix model, as observed in 
\cite{gopakumar}, but this appears to be a coincidence that happens only for
quadratic superpotentials. Whether there is a direct way other than 
(\ref{mres}) to extract the matrix model resolvent remains to be seen.

A further discussion and a proof of relations (\ref{c1})-(\ref{c5}) will be
given in \cite{toappear}.

\section{Interpretation of glueball fields}

Though we have seen that the matrix model resolvent has a direct interpretation
in terms of the Lax matrix, this does not yet explain how to relate it to the
four-dimensional superpotential (\ref{wdv}). One thing that is easy to
do is to figure out what the interpretation of the chiral superfield $S=\sum_i S_i$
is. According to \cite{inin}, the $S$-dependence can be derived from the value
of the superpotential at the minimum by integrating it back in. The integrating
in procedure amounts to replacing $W_{\rm min}(\Lambda)$ by
$W_{\rm min}(\Omega) + S \log(\Lambda^{2N}/\Omega^{2N})$. The value $2N$ in the exponent comes from the coefficient of the $\beta$-function in 3d. 
Integrating out $S$ reproduces $W_{\rm min}(\Omega)$, and the $S$-dependence
is found by integrating out $\Omega$. 

Now the only $\Lambda$-dependence of the effective action is through the
Lagrange multiplier term (\ref{defmult}), and if we integrate out $\Omega$
from
\be
L \log\left( \frac{\Omega^{2N}}{\prod_{i=0}^{N-1} y_i} \right) +
S  \log\left( \frac{\Lambda^{2N}}{\Omega^{2N}} \right)
\ee
we obtain
\be
S   \log\left( \frac{\Lambda^{2N}}{\prod_{i=0}^{N-1} y_i} \right)
\ee 
which is the same as the original Lagrange multiplier term (\ref{defmult}),
except that $L$ has been replaced by $S$. This shows that in general the Lagrange
multiplier $L$ can  be identified with $S=\sum_i S_i$. 
To illustrate this fact, we will next demonstrate how to recover the semiclassical
expansion in four dimensions in terms of $S$ from the three-dimensional superpotential
(\ref{prop}). 
We will consider two cases: $U(2)$ in the maximally confining case and $U(4)$ breaking to $U(2)\times U(2)$. 
After that, we will return to the general case. 

\subsection{$U(2)$}
For convenience we specialize to the superpotential
\begin{equation}
        W(x) = \frac{x^3}{3} - a^2 x,
        \label{eq:supercube}
\end{equation}
which has extrema at $x=\pm a$. Since we are going to study the maximally confining case there is only one chiral superfield $S$, which therefore has to be equal to the Lagrange multiplier $L$:
\begin{displaymath}
        L = S.
\end{displaymath}
As explained above, the $S$-dependence can be recovered from the superpotential
\begin{displaymath}
        W_{\mathrm{eff}} = W(\phi_1)+W(\phi_2) +  (y_0+y_1)(\phi_1+\phi_2) + S \log \left (\frac{\Lambda^4}{y_0 y_1} \right )
\end{displaymath}
by integrating out the $\phi$'s and the $y$'s degrees of freedom. This means we have to solve for the extrema of the superpotential in terms of the $\phi$'s and $y$'s, so we can simply use the results of section 
\ref{sec:u2}\footnote{Since we do not want to integrate out $S$ we have to be careful not to use the equation of motion for $S$: $y_0 y_1 =\Lambda^4$. }. The only thing we have to do is to choose the vacuum. In the classical limit $\phi$ is $\pm a$, we pick $\phi_\mathrm{clas}=a$. To integrate out the $\phi$'s we pick the corresponding solution from section \ref{sec:u2} (with $g_3=1, g_2=-a^2,g_1=0$)
\begin{eqnarray}
        y &=& y_0 = y_1 \nonumber \\ 
        \phi &=& \phi_1=\phi_2 = \sqrt{a^2 - 2 y}.
        \label{eq:semiclassU2sol}
\end{eqnarray}
Plugging this into the superpotential leaves us with
\begin{equation}
        W_{\mathrm{eff}} = -\frac{4}{3} (a^2 - 2 y)^{\frac{3}{2}} + S \log \left (\frac{\Lambda^4}{y^2} \right ).
\end{equation}
The next step is to integrate out the $y$'s
\begin{equation}
        \frac{\partial W}{\partial y} = 0 \Rightarrow S^2= 4 y^2 (a^2 - 2 y)
	\label{eq:yandS}.
\end{equation}
To solve (\ref{eq:yandS}) we write
\begin{displaymath}
	\xi = \frac{S}{4 a^3} \mbox{ and } y(S) = A(\xi) \frac{S}{2a}.
\end{displaymath}
Then (\ref{eq:yandS}) can be written as
\begin{displaymath}
	A^2(\xi)  - 4 \xi A^3(\xi) =1.
\end{displaymath}
In principle there are three solutions for $A(\xi)$, however not all solutions have the right classical limit ($\Lambda \rightarrow 0$). From the solution presented in section \ref{sec:u2} we learn that in the classical limit we need to have $y=\frac{S}{2a}$, therefore:
\begin{displaymath}
	y = \frac{S}{2 a} + \mathcal{O}(S^2) \mbox{, hence } A(0)=1.	
\end{displaymath}
The solution for $A(\xi)$ (with $A(0)=1$) is (see for example \cite{sequence})
\begin{displaymath}
	A(\xi) = \sum_{n=0}^\infty \frac{2^{2n}}{n+1} { \frac{3n-1}{2} \choose n} \xi^n = \frac{1}{12\xi}+\frac{1}{6\xi} \sin\left ( \mathrm{arcsin}\left ( \frac{216 \xi^2 -1}{3} \right )\right)
\end{displaymath}
yielding the following expression for $y$
\begin{displaymath}
\begin{split}
	y(S) &= \frac{S}{2a} \sum_{n=0}^\infty \frac{2^{2n}}{n+1} { \frac{3n-1}{2} \choose n} \left ( \frac{S}{4 a^3} \right )^n = -\frac{a^2}{6}\left ( -1 + 2 \sin\left (\frac{1}{3}\mathrm{arcsin}\left ( 1 - \frac{27 S^2}{2 a^6}\right )\right)\right )\\ &= \frac{S}{2 a} + \frac{S^2}{4 a^4} + \frac{5 S^3}{16 a^7} + \mathcal{O}(S^4).
\end{split}
\end{displaymath}
Substituting this solution into the effective superpotential gives us, in principle, a closed expression valid to all orders in $S$. However the form of this expression is not particularly illuminating and therefore we expand the superpotential in $S$
\begin{equation}
        W_{\mathrm{eff}} = -\frac{4 a^3}{3} + 2 S \left (1-\log \left ( \frac{S}{\Lambda^2 m} \right ) \right ) - \left (\frac{S^2}{2a^3}+\frac{S^3}{3 a^6} + \frac{35 S^4}{96 a^9} + \ldots \right ),
        \label{eq:semiclassU2expansion}
\end{equation}
here $m=2a$ is the mass of the fluctuations of $\phi$ around the classical extremum $\phi=a$.

Let us compare this with the four dimensional answer \cite{dgkv}, in that case we would write (for gauge group $U(N)$)
\begin{displaymath}
        W_{\mathrm{eff}} = N S \left (1 - \log \left (\frac{S}{\Lambda^3}\right ) \right ) - N S \log \left (\frac{\Lambda}{m} \right ) - N \frac{\partial\mathcal{F}}{\partial S}.
\end{displaymath}
For a superpotential $W = \frac{g}{3} \Phi^3 + \frac{m}{2} \Phi^2$ (i.e. $m=2a, g=1$) the function $\mathcal{F}$ is given by
\begin{displaymath}
        \mathcal{F} = \frac{2}{3} \frac{g^2}{m^3} S^3 + \frac{8}{3} \frac{g^3}{m^6} S^4 + \frac{56}{3} \frac{g^4}{m^9} S^5 + \ldots \Rightarrow 2 \frac{\partial\mathcal{F}}{\partial S}= \frac{S^2}{2 a^3} + \frac{S^3}{3 a^6}+\frac{35 S^4}{96 a^9} + \ldots
\end{displaymath}
So we see that equation (\ref{eq:semiclassU2expansion})
 is in good agreement with the four-dimensional answer.

\subsection{$U(4) \rightarrow U(2) \times U(2)$}
In this section we consider the gauge group $U(4)$ and the superpotential as in equation (\ref{eq:supercube}). We expand around the following vacuum
\begin{displaymath}
        \phi_{1,3} =a \qquad \phi_{2,4} = -a.
\end{displaymath}
Classically this vacuum breaks the $U(4)$ to a $U(2)\times U(2)$ symmetry. So in this case there are two chiral superfields involved: $S_1$ and $S_2$. However, we can only integrate in the sum of these, since $L=S=S_1+S_2$. To integrate in $S$ we use the same approach as in the previous section, we integrate out the $\phi$'s and $y$'s.

The first steps are parallel to the calculation done in section~5.
We integrate out the $\phi$'s first, allowing us to express the $\phi$'s in terms of the $y$'s
\begin{displaymath}
        \phi_1 = \sqrt{a^2 - y_0 -y_1} \qquad \phi_3 = \sqrt{a^2-y_2-y_3}
\end{displaymath}
\begin{displaymath}
        \phi_2 = -\sqrt{a^2 - y_1-y_2} \qquad \phi_4 = -\sqrt{a^2 -y_0-y_3}.
\end{displaymath}
We plug this into the superpotential which is then expressed in terms of the $y$'s only. In principle we can proceed to integrate out the $y$'s, however the algebra involved is rather messy, therefore we choose to expand the superpotential as a power series in the $y$'s. Up to second order the superpotential then reads
\begin{displaymath}
        W_\mathrm{eff} = S\log(\frac{\Lambda_4^8}{y_0 y_1 y_2 y_3}) - \frac{1}{m} (y_0-y_2)(y_1-y_3)+\ldots
\end{displaymath}
Integrating out the $y$'s yields
\begin{displaymath}
        y_0  = \frac{S m}{2 y_3},\quad y_1 = -y_3,\quad y_2 = -\frac{ S m}{2 y_3}
\end{displaymath}
and leads to the following effective superpotential
\begin{displaymath}
        W_\mathrm{eff} = S ( 2 + \log(\frac{4 \Lambda_4^8}{m^2 S^2})).
\end{displaymath}
The scale $\Lambda_4$, corresponding to the $U(4)$, can be related to the scales of the $U(2)$'s $\Lambda_2$ (we will write $\Lambda_2^3 = m \Lambda^2 $)
\begin{displaymath}
        \Lambda_2^6 = \frac{\Lambda_4^8}{m^2} = m^2 \Lambda^4 \Rightarrow \Lambda^2_4 = m \Lambda.
\end{displaymath}
So we finally have
\begin{equation}
        W_\mathrm{eff} = 2 S ( 1 - \log(\frac{S}{2 m \Lambda^2})).
        \label{eq:supercubeU4}
\end{equation}

In order to compare this with the four dimensional result \cite{dgkv}
\begin{displaymath}
        W_\mathrm{eff} = 2 ( -S_1 \log(\frac{S_1}{m \Lambda^2}) - S_2 \log(\frac{S_2}{m \Lambda^2}) + S_1+S_2 +\ldots)
\end{displaymath}
we should express this effective superpotential in terms of $S=S_1+S_2$, so we have to integrate out $S_1-S_2$. Since we have expanded the action only to first order in the chiral superfields, the two chiral superfields don't mix and integrating out $S_1-S_2$ is trivial. The result is that $S_1=S_2=S/2$, substituting this in the superpotential gives us back equation (\ref{eq:supercubeU4}).

\subsection{Going from $R^4$ to $R^3 \times S^1$}

Once we identify the Lagrange multiplier $L$ with the glueball field $S=\sum_i S_i$,
there is a concrete procedure to find the superpotential as a function of $S$ from
(\ref{prop}). One may wonder whether one can also go back and start with the
result on $R^4$ and construct the superpotential on $R^3 \times S^1$. We don't know
whether this can be done in general, but a step in this direction is to show
how one can obtain  (\ref{w3}) from (\ref{wvy}). The procedure is very similar
to the path integral derivation of 2d mirror symmetry given in \cite{vafaaganagic}.
Starting with the Veneziano-Yankielowicz superpotential 
\be \label{wvyaa}
W_{VY}(S) = S\left[ \log \left( \frac{\Lambda^{3N}}{S^N} \right) + N \right]
\ee
we first write it in a form as if it were built out of $U(1)$ pieces rather than $U(N)$,
\be \label{wvybb}
W_{VY} \sim \sum_{t=1}^N S_t\left[ \log \left( \frac{\Lambda^{3}}{S_t} \right) + 1 \right] + \sum_{t=1}^{N-1} Z_t(S_t-S_{t+1}).
\ee
If we integrate out the variables $Z_t$, all $S_t$ are identified, and we are back
at the form (\ref{wvyaa}) of the superpotential. However, we proceed by integrating
out the $S_t$ from (\ref{wvybb}) instead. This yields
\be
W\rightarrow \Lambda^3 (e^{Z_1} + e^{Z_2-Z_1} + \ldots + e^{-Z_{N-1}})
\ee
which is indeed precisely of the form (\ref{wvybb}). It would be very interesting
to understand whether and how these transformations can be generalized, perhaps
in the spirit of 2d mirror symmetry, to more complicated situations. 

\subsection{Interpretation of the individual $S_i$}

A full reconstruction of the superpotential (\ref{wdv}) from (\ref{prop})
requires us to not only find the right interpretation of $S=\sum S_i$, but
also of the individual $S_i$. To do this the nature of the semiclassical expansion
discussed in section~5 is probably crucial. If we could write the superpotential
in the form (\ref{semex}) with constraints (\ref{pcons}), we could try to
impose the constraints (\ref{pcons}) using various Lagrange multiplier fields
$L_i$ similar as in (\ref{defmult}), and it would then be natural to identify
those with the $S_i$. To lowest order, this would simply boil down to integrating
in the $S_i$ in each individual gauge group and therefore correctly reproduce
the Veneziano-Yankielowicz superpotentials (\ref{wvy}). We leave a further study
of this to future work.

\section{Conclusions}

In this paper we have described the low-energy effective superpotential
for supersymmetric gauge theories on $R^3\times S^1$. We have shown
that it yields precisely the same vacuum structure as one obtains on $R^4$,
but the relation between the two is highly non-trivial. The results are
a first step towards a direct derivation of the integrable system that
underlies the Dijkgraaf-Vafa matrix model from field theory.

We believe that the formulation of the theory on $R^3 \times S^1$ offers
some advantages over the formulation on $R^4$. For instance, the generalization
of the periodic Toda chain to arbitrary gauge groups is known \cite{swrev2},
and therefore the present formalism should also be applicable to groups like
$G_2$ and $E_6$, for which the Dijkgraaf-Vafa matrix model has not yet been
worked out. We also know the relevant integrable system for various other
gauge theories,as summarized in \cite{swrev3},
such as $N=2$ theories with matter \cite{russians},
$N=4$ super Yang-Mills theory \cite{donagiwitten},
certain conformally invariant $N=2$ theories with gauge groups of quiver type
\cite{donagiwitten,dorey2}, and even for some 5d theories \cite{nekrasov,hollowood}.
Ultimately, we would like to understand in all these cases the nature of the
reduction of the integrable system in a supersymmetric vacuum, thereby generalizing 
the results in section~6. 

In string theory, there is no obvious way to describe 3d field
theory results using topological string theory, since this would require some
7d topological string theory. The integrable system can however in some cases
be extracted
from string theory using dualities and suitable brane configurations 
\cite{kapustin}, and it is worthwhile to explore this connection further. 

The integrable system itself plays a crucial role in this discussion. There are
many physical properties that beg for a nice explanation in terms of the
integrable system, such as for example the loop equations and generalized
Konishi anomaly of \cite{cdsw}. At the same time, there are many features of
the integrable system we have not yet used, such as the description of its
solutions in terms of algebraic-geometric data \cite{alggeom}, and 
such as the existence of
additional flows associated to Whitham times (see e.g. \cite{morerussians} and
references therein). The latter may
help in finding the correct interpretation of the gluino bilinear superfields 
$S_i$.

We found it particularly elegant that lifting vacua from $U(N)$ to $U(tN)$ had such
a nice interpretation in terms of a simple replica trick for the Lax matrix.
In this context it is amusing to observe that there is a close relation between
replica tricks, random matrix theory and the Toda lattice hierarchy 
\cite{condmat0209594}, but whether that is of any relevance to the present discussion
remains to be seen.

\vspace{.5in}

{\bf Acknowledgments: }
We would like to thank Robbert Dijkgraaf, Nick Dorey, Annamaria Sinkovics and
Stefan Vandoren
for useful discussions. This work is partially supported by the Stichting FOM.

\appendix
\section{A recurrence relation for $P_N(x)$}
\label{app:recurrence}
In this appendix we derive a recurrence relation for characteristic polynomials of the following type
\begin{equation}
    P_N(x) = \det(x 1_N -\Phi) =  \begin{vmatrix}
    x-\phi_1 & -y_1 & 0 & . & . & . & 0  & -z\\
    -1 & x-\phi_2 & -y_2 & 0 & . & . & . & 0\\
    0 & -1 & x-\phi_3 & -y_3 & 0 & . & . & 0\\
    . & . & . & . & . & . & . & . \\
    . & . & .& .& .& -1 & x-\phi_{N-2} & -y_{N-1}\\
    -\frac{y_0}{z} & 0 & . & .& .& . & -1 & x-\phi_{N-1}\end{vmatrix}
\end{equation}
This determinant can be expressed in  terms of determinants of the following form
\begin{equation}
    G_N(x) = \begin{vmatrix}
    x-\phi_1 & -y_1 & 0 & . & . & . & 0  & 0\\
    -1 & x-\phi_2 & -y_2 & 0 & . & . & . & 0\\
    0 & -1 & x-\phi_3 & -y_3 & 0 & . & . & 0\\
    . & . & . & . & . & . & . & . \\
    . & . & .& .& .& -1 & x-\phi_{N-2} & -y_{N-1}\\
    0 & 0 & . & .& .& . & -1 & x-\phi_{N-1}\end{vmatrix}.
\end{equation}
Expanding $P_N(x)$ along the bottom line and keeping only z-independent terms we get
\begin{equation}
\begin{split}
    P_N(x) &= (x-\phi_{N-1}) G_{N-1}(x) - (-1)(-y_{N-1}) G_{N-2}(x) + (-1)^{N+1} (-\frac{y_0}{z})(-1)^N(-z) G^+_{N-2}(x) \\ &= (x-\phi_{N-1}) G_{N-1}(x) - y_{N-1} G_{N-2}(x) - y_0 G^+_{N-2}(x),
\end{split}
\end{equation}
here $G^+_N(x)$ is equal to $G_N(x)$ with shifted $\phi$'s and y's (i.e. $\phi_i \rightarrow \phi_{i+1}, y_i \rightarrow y_{i+1}$).

The $G_N(x)$ and $G^+_N(x)$ are tri-diagonal and therefore satisfy the recurrence relations
\begin{equation}
    G_N(x) = (x-\phi_{N-1}) G_{N-1}(x) - y_{N-1} G_{N-2}(x)
\end{equation}
\begin{equation}
    G^+_N(x) = (x-\phi_N) G^+_{N-1}(x) - y_N G^+_{N-2}(x).
\end{equation}

As a special case, take all the y's and $\phi$'s equal, then $G^+_N(x)=G_N(x)$ and the recurrence relation for $P_N(x)$ is
\begin{displaymath}
    P_N(x) = (x-\phi) G_{N-1} - 2 y G_{N-2} = G_N(x) - y G_{N-2},
\end{displaymath}
in this case it is easy to show that the $P_N$ satisfies the same recurrence relation as the $G_N$
\begin{equation}
    P_N(x) = (x-\phi) P_{N-1}(x) - y P_{N-2}(x),
    \label{eq:recur_chebyshev}
\end{equation}
which is, up to some rescaling, the Chebyshev recurrence relation. Therefore the $P_N(x)$ (with all y's and $\phi$'s equal) are given by the Chebyshev polynomials of the first or of the second kind.


\end{document}